\documentclass[preprint,3p,12pt,sort&compress]{elsarticle}

\usepackage{amssymb}
\usepackage{lineno}
\usepackage[colorlinks,linkcolor=blue,anchorcolor=blue,citecolor=blue,bookmarksnumbered,bookmarksopen]{hyperref}
\usepackage{amsmath}
\usepackage{bm}
\usepackage{caption}

\usepackage{multirow}
\usepackage{booktabs}
\usepackage{float}
\usepackage{subfigure}

\journal{Thin-Walled Structures}

\begin{document}
\captionsetup[figure]{labelfont={bf},name={Fig.},labelsep=period}
\captionsetup[table]{labelfont={bf},labelsep=newline,singlelinecheck=false}

\begin{frontmatter}



\title{A semi-analytical method using auxiliary sine series for vibration and sound radiation of a rectangular plate with elastic edges}


\author[1]{Guoming Deng}
\ead{dengguoming@outlook.com}

\author[2]{Xian Wu}

\author[3]{Changxiao Shao}

\author[2]{Songlin Zheng}

\author[2]{Jianwang Shao\corref{cor1}}
\ead{shaojianwang@tongji.edu.cn}

\cortext[cor1]{Corresponding author}
%%\cortext[cor2]{corresponding author}

\address[1]{Shenzhen Technology University, Shenzhen 518118, PR China}
\address[2]{School of Automotive Studies, Tongji University, Shanghai 201804, PR China}
\address[3]{School of Mechanical Engineering and Automation, Harbin Institute of Technology (Shenzhen), Shenzhen 518055, PR China}

\begin{abstract}
This paper proposes an efficient semi-analytical method using auxiliary sine series for transverse vibration and sound radiation of a thin rectangular plate with edges elastically restrained against translation and rotation. The formulation, constructed by two-dimensional sine and/or cosine series, can approximately express the bending displacement, and calculate vibration and sound radiation under excitation of point force, arbitrary-angle plane wave, or diffuse acoustic field with acceptable accuracy. It is also applied for baffled or unbaffled conditions. A post-process program is developed to predict vibrating frequencies and modes, mean square velocity spectrum, and sound transmission loss via reduced-order integrals of radiation impedances. The method is validated by experiment and simulation results, demonstrating accurate and efficient computation using a single program for transverse vibration and sound radiation of a plate under different elastic boundary conditions and different excitations. Formulas given in this paper provide a basis for the code development on transverse vibration and sound radiation analysis of thin plates.
\end{abstract}


%%\item An added sine-auxiliary function for the Fourier series expression of cavity modes
%%\item Numerical modeling on the structural-acoustic coupling problem
%%\item Optimization on sectional geometry with a modified simulated annealing algorithm
%%\item Improved performance on the sound insulation of basic bulb seal models
%%\addcontentsline{toc}{section}{Highlights}


\end{frontmatter}

%%\linenumbers

%%---------------------------%%%%%%%%%%%%%%%%Section 1%%%%%%%%%%%%%%%
\section{Introduction}
\label{sec:intro}
The vibration and sound radiation of thin plates under mechanical \cite{Kim2020, Shao2021, Shao2023}, acoustic \cite{Marchetto2017, Deng2021a}, and/or random excitations \cite{Bonness2017, Hambric2004, Shao2020} are very common in aeronautical, automotive, and marine fields. Aircraft fuselage panel \cite{Hasan2023, Ghiringhelli2013}, automotive panel and glass \cite{Shao2020, Shao2021, He2021}, ship hull and deck \cite{Thekinen201973} are typical thin plate structures, whose bending vibration and sound radiation are important research topics for reducing airborne noise. Although the finite element method (FEM), boundary element method (BEM), and hybrid FE-SEA (statistical energy analysis) are used for numerical analysis of vibration and noise in complex structures \cite{Sehgal2016, CHEN201422, ZHANG2023110630}, due to the computational capability and cost, such methods have difficulties in design changes and optimization of noise reduction. Simplified approaches or semi-analytical methods provide fast and low-cost means for parameter studies and design optimizations \cite{Deng2020, Deng2021b, Liu2024, Liu2023a, Liu2023b}.

Early investigations included an exact solution of the differential equation of a plate with bending vibration and the numerical procedure for analytical calculations of vibrating frequencies and modes. The possible combinations of classical boundary conditions (i.e., simply supported, clamped and free) of rectangular plates produce 21 distinct cases \cite{Leissa1973}. The first exact solution was obtained for the case having four edges simply supported \cite{Navier1819}, which can be found in the books of Rayleigh \cite{Rayleigh1945} and Timoshenko \cite{Timoshenko1934}. Five more cases were solved by Levy \cite{Levy1899}. The well-known exact solutions of the six cases having one pair of opposite edges simply supported and any conditions at the other two edges were presented by Voigt \cite{Voigt1893}. Around 1909, Ritz \cite{Ritz1909} developed a numerical procedure to approximately calculate the frequencies and modes of a rectangular plate with all four edges free by extending the Rayleigh principle. Pickett and Kan \cite{Pickett1939} followed this method, which was also called generalized energy method, to find solutions of a rectangular clamped plate. Similarly, Young \cite{Young1950} used the method and called it Ritz method, to obtain solutions of three specific plate problems, i.e., rectangular plate clamped at all four edges, rectangular plate clamped along two adjacent edges and free along the other two edges, and rectangular plate clamped along one edge and free along the other three edges. According to Rayleigh principle, the frequencies calculated by Ritz's procedure are always higher than the exact values, because the assumption of an incorrect waveform for the vibrating displacement of a rectangular plate introduces constraints to the system. Warburton \cite{Warburton1954} called it the Rayleigh-Ritz method and used the procedure to derive simple approximate frequency expressions for 15 cases of combined boundary conditions. The assumed waveforms were like those of beams and the assumed functions were called beam functions by other scholars \cite{Leissa1973, DICKINSON19781}.

Afterwards, the flexural vibrations of rectangular plates with edges elastically restrained against translation and rotation were investigated. The classical boundary conditions are the special cases of the situation, where eight different translational and rotational flexibility coefficients are considered for the four edges. Laura and Grossi \cite{LAURA1981101} calculated the frequencies of a rectangular plate with elastically restrained edges by using Rayleigh-Ritz method and polynomials as the assumed functions. Warburton and Edney \cite{WARBURTON1984537} determined the frequencies by using Rayleigh-Ritz method and assumed functions of beams. Whether the assumed functions were polynomials or beam functions, acceptable accuracy could be obtained. Orthogonal polynomials were later employed for obtaining natural frequency and/or buckling loads of rectangular plates based on Rayleigh-Ritz method \cite{BHAT1985493, CUPIAL1997385, DICKINSON198651}. Berry et al. \cite{Berry1990} used simple polynomials as the trial (or assumed) functions and Rayleigh-Ritz method to calculate sound radiation from rectangular, baffled plates with elastic boundary conditions. It was pointed out that the selection of polynomial functions had advantages in obtaining both resistive and reactive components of radiation impedance of the plate and furnishing a natural decomposition of the radiator into monopoles, dipoles, etc. \cite{Berry1994}. The general elastic boundary conditions may be extended to viscoelastic supports by assigning the translational and rotational springs complex stiffness, as the work by Park et al. \cite{Park2003, Park2004}, who used the modified beam functions as the trial functions and developed an analytical model for the flow-induced vibration of rectangular plates with viscoelastic supports. In short, the beam functions contain trigonometric function, hyperbolic function, polynomial function, and exponential function. Trigonometric and hyperbolic functions are deployed in a fixed beam; polynomial function appears in free boundary condition \cite{Warburton1954}. Exponential function was involved in the modified beam function by Park et al. \cite{Park2003}. For beams with different boundary conditions, like fixed-fixed, fixed-free, free-free, etc., different combinations of these functions need to be used. On the other hand, Beslin and Nicolas \cite{Beslin1997} showed that using polynomials as trial functions often resulted in ill-conditioned mass and stiffness matrices due to numerical roundoff errors when predicting very high order modes.

Recently, more attention has been paid to beam functions, especially its embedded trigonometric function when studying the bending vibration of thin plates with general elastic boundary condition. Li et al. \cite{Li2009} used a two-dimensional Fourier series supplemented with a few one-dimensional Fourier series to express the vibrating displacement of a plate. The auxiliary functions were chosen in the form of special trigonometric functions, which ensured that both the governing equation and boundary conditions were satisfied exactly on a point-wise basis. Various boundary conditions were checked for the frequencies and modes. Following this approach, Mohammadesmaeili et al. \cite{MOHAMMADESMAEILI2021104274} used auxiliary polynomial functions to construct Fourier series expansions for free vibration and buckling analysis of a rectangular Mindlin plate. Different supplementary polynomials were employed by Wu et al. \cite{Wu2022}. Modified Fourier series methods were proposed for other applications, e.g., free vibration of the moderately thick laminated composite rectangular plate \cite{ZHANG20171}, vibration of rotating toroidal shell \cite{SENJANOVIC2018870}, displacement solution for isotropic linear elastic solids under plane strain or plane stress states \cite{Barulich2022}, deformation and stress solutions of rectangular anisotropic plates \cite{Xu2022}. Trigonometric functions were assumed for the displacement or stress solutions, showing the powerfulness of Fourier series in expressing general elastic boundary conditions. However, few investigations have focused on the sound radiation of plates by using the Fourier series methods.

Sound radiation problem needs to solve a quadruple integral of the radiation impedance. Nelisse et al. \cite{Nelisse1998} used a trigonometric function set for the basis or assumed functions in expressing vibration of a rectangular plate. As a result, the double integral could be turned into single one. The radiation efficiency of a clamped plate in water was accurately predicted by their method up to 500 Hz. Zhang and Li \cite{ZHANG20105307} calculated the sound radiation by converting the surface integral into simple integral based on MacLaurin series expansion. However, the prediction at high frequency was not so effective. Zhang et al. \cite{Zhang202313} used a modified supplemented two-dimensional Fourier cosine series following the work of Zhang and Li \cite{ZHANG20105307}. Zhao et al. \cite{Zhao2024} proposed a fast Chebyshev spectral approach to solve the vibroacoustic response of heavy fluid-loaded baffled rectangular plates. The aforementioned research used different combinations of trigonometric functions or polynomials as assumed functions and did not use auxiliary sine series which are favorable for calculation of radiation impedance matrices and high-frequency solutions.

In this paper, a semi-analytical method using Fourier sine series as the auxiliary function for the first time is proposed to predict transverse vibration and sound transmission loss (TL) of a rectangular plate with arbitrarily elastic edges. Very few work investigated analytical or semi-anlytical approaches solving sound insulation problems under diffuse acoustic field (DAF) regarding high frequencies. The currently proposed method may efficiently compute vibrating response, sound radiation, and TLs under different excitations (e.g., single-point force, a plane wave, DAF) and different conditions (e.g., baffled or unbaffled) when involving high-frequency modes. Formulas are derived for developing a post-process Python package. The semi-analytical method and formulations are presented in Section \ref{sec:Theor}, followed by derived formulas based on the auxiliary sine series in Section \ref{sec:Derived_formulas}. Results of cases including multiple excitations, baffled/unbaffled conditions, and TLs of different elastic boundary conditions are discussed in Section \ref{sec:Results}. This work is concluded in Section  \ref{sec:conclusions}.

%%---------------------------%%%%%%%%%%%%%%%%Section 2%%%%%%%%%%%%%%%%%---------------------------------
\section{Semi-analytical method and formulations}
\label{sec:Theor} 

This section gives the semi-analytical method and formulations, which are employed to predict the bending vibration and sound radiation of a thin rectangular plate with elastic edges. The vibration modeling is firstly presented in Section \ref{subsec:VibrationPlate}. Then, the sound radiation impedances of a baffled and unbaffled plate are presented in Section \ref{subsec:SR_baffled} and \ref{subsec:SR_unbaffled}, respectively.

%%---------------------------%%%%%%%%%%%%%%%%Subsection 2.1%%%%%%%%%%%%%%%
\subsection{Vibration of a rectangular plate}
\label{subsec:VibrationPlate}

\begin{figure}
\centering
\includegraphics[scale=.16]{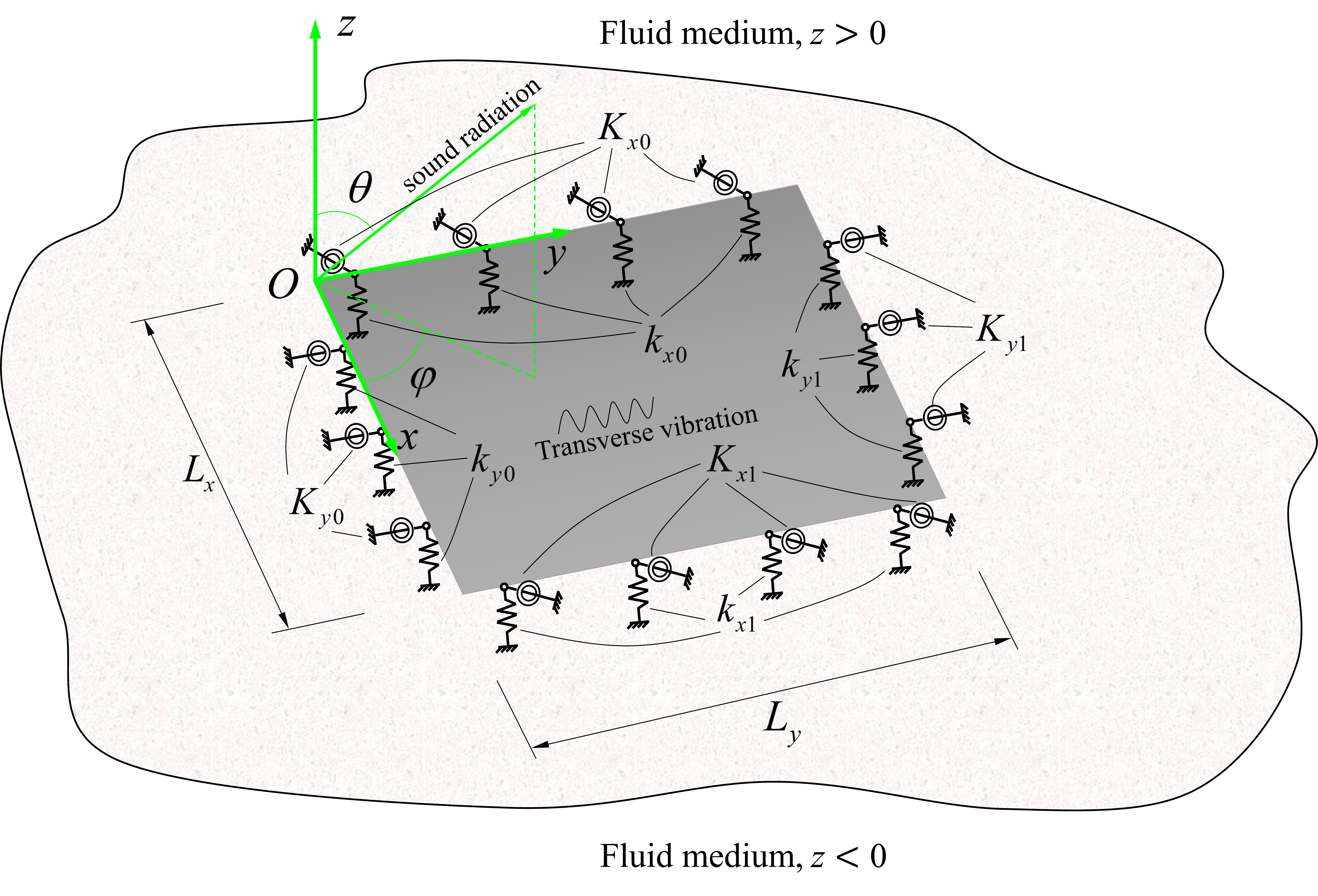}
\caption{Graphical description of a baffled plate with arbitrarily-elastic edges.}
\label{fig:PlateGraph}
\end{figure}
The bending motion of a rectangular plate can be described by a variational formulation. The Hamiltonian for the vibrational plate is obtained through the action integral of energy and work between two arbitrary times \cite{Berry1994,Nelisse1998}. The Rayleigh-Ritz method is used to find the approximate extremum of the Hamiltonian. This study aims to propose a new displacement formulation for a rectangular plate with arbitrarily-elastic edges (see Fig. \ref{fig:PlateGraph}). Assuming a harmonic state $\exp({\rm j}\omega t)$ for the time dependence of the motion, the bending displacement of the plate, $w(x,y)$, is expressed as

\begin{equation}\label{eq:DisSeries}
\begin{aligned}
w(x,y)&=\sum_{m=0}^{\infty}\sum_{n=0}^{\infty}a_{mn}\cos(\lambda_m^{L_x}x)\cos(\lambda_n^{L_y}y)+\sum_{m=0}^{\infty}\sum_{n=0}^{\infty}b_{mn}\cos(\lambda_m^{L_x}x)\sin(\lambda_n^{L_y}y) \\
&+\sum_{m=0}^{\infty}\sum_{n=0}^{\infty}c_{mn}\sin(\lambda_m^{L_x}x)\cos(\lambda_n^{L_y}y),
\end{aligned}
\end{equation}
where $\lambda_m^{L_x}=m\pi/L_x$, $\lambda_n^{L_y}=n\pi/L_y$. The $L_x$ and $L_y$ represent the length and width of the rectangular plate, respectively. The terms $a_{mn}$, $b_{mn}$, $c_{mn}$ indicate amplitudes in the generalized coordinates, and the subscript $m$ or $n$ starts from 0. A singular value decomposition algorithm will be chosen to solve the linear equations derived later.

The sine series expansions in the second and third items on the right side of Eq. (\ref{eq:DisSeries}) can be regarded as the auxiliary coefficients of the one-dimensional cosine series. As a result, we obtain a two-dimensional trigonometric series for either item. They are added to the two-dimensional cosine series (see the first item). The added items tackle discontinuities at the two sets of opposite edges (in x- and y-direction, respectively) due to spatial partial derivatives of the admissible functions. Li et al. \cite{Li2009} designed four specific trigonometric functions as the auxiliary coefficients of two one-dimensional cosine series to address the discontinuities at the edges. Zhang et al. \cite{Zhang2017} deployed six auxiliary sine functions. In our work, the constructed expression can be regarded as a combination of sine and cosine series. This unified form of two-dimensional trigonometric series, i.e., Eq. (\ref{eq:DisSeries}), is conducive to the derivation of mass and stiffness matrices, as will be seen in Section \ref{subsec:CMSM}.

The displacement of bending plate converges with number of series terms \cite{Deng2020,Li2009,Wang2016-2}. Provided that the expression in Eq. (\ref{eq:DisSeries}) took $M$ and $N$ terms for x- and y-dimension, respectively, the expression is rewritten in a column-vector form,

\begin{equation}\label{eq:DisVector}
w(x,y)=\mathbf{\Gamma}^T(x,y)\mathbf{a},
\end{equation}
where the superscript $T$ indicates the transpose of vector or matrix. $\mathbf{\Gamma}(x,y)$ and $\mathbf{a}$ are column vectors of mode function and amplitudes in modal coordinates for the plate, respectively. The detailed expressions for $\mathbf{\Gamma}(x,y)$ and $\mathbf{a}$ are provided in \ref{sec:A1}. 

The stationarity condition for the Hamiltonian can be expressed by Lagrange's equation of motion \cite{Stupakov2018,Li2004}. The expressions of total potential energy and kinetic energy can be referred to a previous report \cite{Deng2020b}. Solving the Lagrange's equation, we obtain the characteristic equation for the free vibration of the plate as

\begin{equation}\label{eq:PlateMatixEq}
\Bigl(\mathbf{K}-\omega^2\mathbf{M}\Bigr)\mathbf{a}=\mathbf{0}.
\end{equation}
The stiffness matrix $\mathbf{K}$ and the mass matrix $\mathbf{M}$ are given by

\begin{equation}\label{eq:K_Matrix}
\begin{aligned}
\mathbf{K}&=D\int_0^{L_x}\int_0^{L_y}\Bigl(\mathbf{\Gamma}_{xx}\mathbf{\Gamma}_{xx}^T+\mathbf{\Gamma}_{yy}\mathbf{\Gamma}_{yy}^T+\mu\mathbf{\Gamma}_{xx}\mathbf{\Gamma}_{yy}^T+\mu\mathbf{\Gamma}_{yy}\mathbf{\Gamma}_{xx}^T+2(1-\mu)\mathbf{\Gamma}_{xy}\mathbf{\Gamma}_{xy}^T\Bigr){\rm d}y{\rm d}x \\
&+\int_0^{L_y}\Bigl(k_{x0}\mathbf{\Gamma}\mathbf{\Gamma}^T+K_{x0}\mathbf{\Gamma}_{x}\mathbf{\Gamma}_{x}^T\Bigr)_{x=0}{\rm d}y + \int_0^{L_y}\Bigl(k_{x1}\mathbf{\Gamma}\mathbf{\Gamma}^T+K_{x1}\mathbf{\Gamma}_{x}\mathbf{\Gamma}_{x}^T\Bigr)_{x=L_x}{\rm d}y \\
&+\int_0^{L_x}\Bigl(k_{y0}\mathbf{\Gamma}\mathbf{\Gamma}^T+K_{y0}\mathbf{\Gamma}_{y}\mathbf{\Gamma}_{y}^T\Bigr)_{y=0}{\rm d}x + \int_0^{L_x}\Bigl(k_{y1}\mathbf{\Gamma}\mathbf{\Gamma}^T+K_{y1}\mathbf{\Gamma}_{y}\mathbf{\Gamma}_{y}^T\Bigr)_{y=L_y}{\rm d}x,
\end{aligned}
\end{equation}

\begin{equation}\label{eq:M_Matrix}
\begin{aligned}
\mathbf{M}=\rho h\int_0^{L_x}\int_0^{L_y}(\mathbf{\Gamma}\mathbf{\Gamma}^T){\rm d}y{\rm d}x,
\end{aligned}
\end{equation}
where $\mathbf{\Gamma}_{xx}$ is the second-order partial derivative of $x$ for $\mathbf{\Gamma}$, i.e., $\mathbf{\Gamma}_{xx}=\partial\mathbf{\Gamma}_x/\partial x=\partial^2\mathbf{\Gamma}/\partial x^2$. Analogously, other second-order derivatives include terms $\mathbf{\Gamma}_{yy}$, $\mathbf{\Gamma}_{xy}$. $D=Eh^3/\bigl[12(1-\mu^2)\bigr]$ is the bending stiffness with $E$ and $\mu$ indicating the elastic modulus and Poisson’s ratio, respectively. The terms $\rho$, $h$, $\omega$ indicate the bulk density, thickness of the plate, and the circular frequency, respectively. Concerning the damping loss factor $\eta$, the complex value of the elastic modulus is given by $E=E_0(1+{\rm j}\eta)$, where $E_0$ is Young's modulus of the isotropic plate. The terms $k_{x0}$, $k_{x1}$ ($k_{y0}$, $k_{y1}$) and $K_{x0}$, $K_{x1}$ ($K_{y0}$, $K_{y1}$) indicate the transverse and rotational spring constants at $x=0$, $x=L_x$ ($y=0$, $y=L_y$), respectively, seen in Fig. \ref{fig:PlateGraph}. The semi-analytical model could be extended to the case of stiffened plate by adding stiffener's energy according to first-order shear deformation theory and geometrical relations, as in the work by Gao et al. \cite{Gao2022}, who used Jacobi orthogonal polynomials and trigonometric series.

%%---------------------------%%%%%%%%%%%%%%%%Subsection 2.2%%%%%%%%%%%%%%%
\subsection{Sound radiation of a baffled plate}
\label{subsec:SR_baffled}

This subsection gives the sound radiation modeling of a baffled plate based on the proposed bending displacement formulation. The acoustic pressure load on the plate due to its vibration can be expressed by using a sound radiation impedance matrix \cite{Nelisse1998}. Considering fluid medium on both sides of a baffled plate (see Fig. \ref{fig:PlateGraph}), we have the linear equations under an external force as

\begin{equation}\label{eq:MotionEq_baffled}
\bigl(\mathbf{K}-\omega^2\mathbf{M}+2{\rm j}\omega\mathbf{Z}_{\rm bf}\bigr)\mathbf{a}=\mathbf{f},
\end{equation}
where $\mathbf{f}$ is the modal force that will be given later. The radiation impedance matrix for a baffled plate $\mathbf{Z}_{\rm bf}$ is obtained through the calculation of Rayleigh integral, 

\begin{equation}\label{eq:RadiationMatrix_baffled}
\mathbf{Z}_{\rm bf} = {\rm j}\rho_0\omega\int_0^{L_x}\int_0^{L_y}\int_0^{L_x}\int_0^{L_y}\bigl[\mathbf{\Gamma}(x,y)G(x,y,0;x',y',0)\mathbf{\Gamma}^T(x',y')\bigr]{\rm d}y{\rm d}x{\rm d}y'{\rm d}x',
\end{equation}
where $\rho_0$ is the density of the fluid medium, and $G(x,y,0;x',y',0)={\rm exp}(-{\rm j}k_0 R)/(2\pi R)$ is the Green function, where $R=\sqrt{(x-x')^2+(y-y')^2}$ and $k_0=\omega/c_0$. The $c_0$ indicates the sound speed in the medium. 

%%---------------------------%%%%%%%%%%%%%%%%Subsection 2.3%%%%%%%%%%%%%%%
\subsection{Sound radiation of an unbaffled plate}
\label{subsec:SR_unbaffled}

According to the derivation from Nelisse et al. \cite{Nelisse1998}, who used an expression over admissible functions to identify the pressure jump, the linear equations for the unbaffled plate under an external force are given as

\begin{equation}\label{eq:MotionEq_unbaffled}
\begin{aligned}
\begin{bmatrix}\mathbf{K}-\omega^2\mathbf{M} & \mathbf{E}^T \\
\mathbf{E} & \mathbf{F}/\omega^2 \\ \end{bmatrix} \begin{bmatrix}\mathbf{a} \\ \tilde{\mathbf p} \\ \end{bmatrix} = \begin{bmatrix}\mathbf{f} \\ \mathbf{0} \\ \end{bmatrix},
\end{aligned}
\end{equation}
where the column vector $\tilde{\mathbf p}$ indicates the amplitudes of the generalized pressure jump. A set of admissible functions used for a simply-supported plate, i.e., $\mathbf\Lambda$ with entries $\mit\Lambda_{mn}={\rm sin}(\lambda_m^{L_x}x){\rm sin}(\lambda_n^{L_y}y)$, are employed here to identify the pressure jump due to the vanishing characteristics of the sine function at the boundary. For the sake of unity, the subscripts $m,n$ start from 0. The pressure jump is then mapped to the $\mathbf\Lambda$ basis, i.e., $f_P(x,y)=\mathbf{\Lambda}^T(x,y)\tilde{\mathbf p}$. The matrix $\mathbf{E}$ gives a transform from a $\mathbf\Gamma$ basis to a ${\mathbf\Lambda}$ basis, i.e., $\mathbf{E}=\iint_{S_0} {\mathbf\Lambda}\mathbf{\Gamma}^T {\rm d}S$. The matrix $\mathbf{F}$ behaves like an admittance, i.e.,

\begin{equation}\label{eq:Matrix_F}
\mathbf{F} = -\omega^2\hat{\mathbf{M}}+\hat{\mathbf{K}},
\end{equation}
with

\begin{equation}\label{eq:Matrix_Mhat}
\hat{\mathbf{M}} = \frac{1}{2\rho_0 c_0^2}\int_0^{L_x}\int_0^{L_y}\int_0^{L_x}\int_0^{L_y}\Bigl[{\mathbf\Lambda}(x,y)G(x,y,0;x',y',0){\mathbf\Lambda}^T(x',y')\Bigr]{\rm d}y{\rm d}x{\rm d}y'{\rm d}x',
\end{equation}

\begin{equation}\label{eq:Matrix_Khat}
\begin{aligned}
\hat{\mathbf{K}} = \frac{1}{2\rho_0}\int_0^{L_x}\int_0^{L_y}\int_0^{L_x}\int_0^{L_y} \Bigl[ & {\mathbf\Lambda}_{x}(x,y)G(x,y,0;x',y',0){\mathbf\Lambda}_{x'}^T(x',y') \\ 
& {\mathbf\Lambda}_{y}(x,y)G(x,y,0;x',y',0){\mathbf\Lambda}_{y'}^T(x',y') \Bigr]{\rm d}y{\rm d}x{\rm d}y'{\rm d}x',
\end{aligned}
\end{equation}
where ${\mathbf\Lambda}_x$ and ${\mathbf\Lambda}_y$ are the first-order partial derivatives of ${\mathbf\Lambda}$ with respect to $x$ and $y$.

According to Eq. (\ref{eq:MotionEq_unbaffled}), the radiation impedance matrix for an unbaffled plate is obtained, i.e., ${\mathbf{Z}_{\rm ubf}} = {\rm j}\omega \mathbf{E}^T\mathbf{F}^{-1}\mathbf{E}$, with $\mathbf{E}^T\tilde{\mathbf{p}}={\rm j}\omega{\mathbf{Z}_{\rm ubf}}\mathbf{a}$. However, the computation of the inverse of $\mathbf{F}$ should be avoided. Thus, Eq. (\ref{eq:MotionEq_unbaffled}) is preferred for calculation.

%%---------------------------%%%%%%%%%%%%%%%Section 3%%%%%%%%%%%%%%%%%%---------------------------------
\section{Derived formulas}
\label{sec:Derived_formulas} 

This section presents derived formulas for calculating mass and stiffness matrices, radiation impedance matrices under baffled/unbaffled conditions, and modal forces of point excitation, obliquely incident plane wave, and DAF excitation.

%%---------------------------%%%%%%%%%%%%%%%%Subsection 3.1%%%%%%%%%%%%%%%
\subsection{Calculation of mass and stiffness matrices}
\label{subsec:CMSM}

By using the Fourier sine series expansions as the auxiliary functions, we can acquire simple relational equations for the calculation of mass and stiffness matrices. Four constant vectors are firstly defined as follows: $\mathbf{q}_M^{}$ and $\mathbf{q}_N^{}$ contain entries equal to $(\lambda_m^{L_x})^2, m=0,1,2,\cdots,M$ and $(\lambda_n^{L_y})^2, n=0,1,2,\cdots,N$, respectively; $\mathbf{1}_M^{}$ and $\mathbf{1}_N^{}$ contain $M+1$ and $N+1$ entries all equal to one, respectively.

The mode function vector $\mathbf{\Gamma}(x,y)$ (see \ref{eq:A-1}) is rewritten as the block vectors, $\mathbf{\Gamma}=[\mathbf{\Gamma}_1^T,\mathbf{\Gamma}_2^T,\mathbf{\Gamma}_3^T]^T$, corresponding to the three items on the right side of Eq. (\ref{eq:DisSeries}). Then, we can derive the following formulations,

\begin{equation}\label{eq:Gammaxx}
\mathbf{\Gamma}_{xx}=\partial^2\mathbf{\Gamma}/\partial x^2=\Bigl[\bigl(\bm{\alpha}_1\ast\mathbf{\Gamma}_1\bigr)^T,\bigl(\bm{\alpha}_1\ast\mathbf{\Gamma}_2\bigr)^T,\bigl(\bm{\alpha}_1\ast\mathbf{\Gamma}_3\bigr)^T\Bigr]^T,
\end{equation}

\begin{equation}\label{eq:Gammayy}
\mathbf{\Gamma}_{yy}=\partial^2\mathbf{\Gamma}/\partial y^2=\Bigl[\bigl(\bm{\alpha}_2\ast\mathbf{\Gamma}_1\bigr)^T,\bigl(\bm{\alpha}_2\ast\mathbf{\Gamma}_2\bigr)^T,\bigl(\bm{\alpha}_2\ast\mathbf{\Gamma}_3\bigr)^T\Bigr]^T,
\end{equation}

\begin{equation}\label{eq:Gammaxy}
\begin{aligned}
\mathbf{\Gamma}_{xy}=\partial^2\mathbf{\Gamma}/(\partial x \partial y)=\Biggl\{&\Bigl\{\lambda_m^{L_x}\lambda_n^{L_y}{\rm sin}(\lambda_m^{L_x}x){\rm sin}(\lambda_n^{L_y}y)\Bigr\}_{1,(M+1)(N+1)}^T, \\
&\Bigl\{-\lambda_m^{L_x}\lambda_n^{L_y}{\rm sin}(\lambda_m^{L_x}x){\rm cos}(\lambda_n^{L_y}y)\Bigr\}_{1,(M+1)(N+1)}^T, \\
&\Bigl\{-\lambda_m^{L_x}\lambda_n^{L_y}{\rm cos}(\lambda_m^{L_x}x){\rm sin}(\lambda_n^{L_y}y)\Bigr\}_{1,(M+1)(N+1)}^T \Biggr\}^T,
\end{aligned}
\end{equation}
where ${\bm\alpha_1}={\rm vec}(-\mathbf{1}_N^{}\mathbf{q}_M^T)$, ${\bm\alpha}_2={\rm vec}(-\mathbf{q}_N^{}\mathbf{1}_M^T)$. The function $\rm vec()$ gives vectorization of matrices and the symbol $\ast$ indicates the Hadamard product of two matrices or vectors. Therefore, matrices from the tensor product of the second-order partial derivatives of the mode function vector $\bm\Gamma$ can be calculated. 

According to the block form of $\bm\Gamma$, the matrices ${\bm\Gamma}{\bm\Gamma}^T$, $\bm\Gamma_{xx}\bm\Gamma_{xx}^T$, $\bm\Gamma_{yy}\bm\Gamma_{yy}^T$, $\bm\Gamma_{xx}\bm\Gamma_{yy}^T$, and $\bm\Gamma_{yy}\bm\Gamma_{xx}^T$ can be written as $3\times3$ matrix blocks. Taking the first one of the total nine matrix blocks as an example, we have the following formulas,

\begin{equation}\label{eq:MatrixGammaxxxx}
\begin{aligned}
\bigl[\mathbf{\Gamma}_{xx}\mathbf{\Gamma}_{xx}^T\bigr]_{11}&=\Bigl(\bm{\alpha}_1\ast\mathbf{\Gamma}_1\Bigr)\Bigl(\bm{\alpha}_1\ast\mathbf{\Gamma}_1\Bigr)^T \\
&=\Bigl(\bigl(\mathbf{q}_M^{}\otimes\mathbf{1}_N^{}\bigr)\bigl(\mathbf{q}_M^{}\otimes\mathbf{1}_N^{}\bigr)^T\Bigr)\ast\bigl(\mathbf{\Gamma}_1\mathbf{\Gamma}_1^T\bigr) \\
&=\bigl[\bigl(\mathbf{q}_M^{}\mathbf{q}_M^T\bigr)\otimes\bigl(\mathbf{1}_N^{}\mathbf{1}_N^T\bigr)\bigr] \ast \bigl[\mathbf{\Gamma}\mathbf{\Gamma}^T\bigr]_{11},
\end{aligned}
\end{equation}

\begin{equation}\label{eq:MatrixGammayyyy}
\begin{aligned}
\bigl[\mathbf{\Gamma}_{yy}\mathbf{\Gamma}_{yy}^T\bigr]_{11}&=\Bigl(\bm{\alpha}_2\ast\mathbf{\Gamma}_1\Bigr) \Bigl(\bm{\alpha}_2\ast\mathbf{\Gamma}_1\Bigr)^T \\
&=\bigl[\bigl(\mathbf{1}_M^{}\mathbf{1}_M^T\bigr)\otimes\bigl(\mathbf{q}_N^{}\mathbf{q}_N^T\bigr)\bigr] \ast \bigl[\mathbf{\Gamma}\mathbf{\Gamma}^T\bigr]_{11},
\end{aligned}
\end{equation}

\begin{equation}\label{eq:MatrixGammaxxyy}
\begin{aligned}
\bigl[\mathbf{\Gamma}_{xx}\mathbf{\Gamma}_{yy}^T\bigr]_{11}=\bigl[\bigl(\mathbf{q}_M^{}\mathbf{1}_M^T\bigr)\otimes\bigl(\mathbf{1}_N^{}\mathbf{q}_N^T\bigr)\bigr] \ast \bigl[\mathbf{\Gamma}\mathbf{\Gamma}^T\bigr]_{11},
\end{aligned}
\end{equation}

\begin{equation}\label{eq:MatrixGammayyxx}
\begin{aligned}
\bigl[\mathbf{\Gamma}_{yy}\mathbf{\Gamma}_{xx}^T\bigr]_{11}=\bigl[\bigl(\mathbf{1}_M^{}\mathbf{q}_M^T\bigr)\otimes\bigl(\mathbf{q}_N^{}\mathbf{1}_N^T\bigr)\bigr] \ast \bigl[\mathbf{\Gamma}\mathbf{\Gamma}^T\bigr]_{11},
\end{aligned}
\end{equation}
where the symbol $\otimes$ indicates the Kronecker product of two matrices or vectors. Analogously, the other eight matrix blocks can be obtained by using the corresponding coefficient matrices. 

The following two relations should be employed for the derivation of ${\mathbf\Gamma}_{xy}{\bm\Gamma}_{xy}^T$,

\begin{equation}\label{eq:Integral1}
\bigl(\lambda_s^{L_x}\bigr)^2\int_0^{L_x}{\rm sin}\bigl(\lambda_m^{L_x}x\bigr){\rm cos}\bigl(\lambda_s^{L_x}x\bigr){\rm d}x = -\lambda_m^{L_x}\lambda_s^{L_x}\int_0^{L_x}{\rm sin}\bigl(\lambda_s^{L_x}x\bigr){\rm cos}\bigl(\lambda_m^{L_x}x\bigr){\rm d}x,
\end{equation}

\begin{equation}\label{eq:Integral2}
\begin{aligned}
\bigl(\lambda_m^{L_x}\bigr)^2\int_0^{L_x}{\rm sin}\bigl(\lambda_m^{L_x}x\bigr){\rm cos}\bigl(\lambda_s^{L_x}x\bigr){\rm d}x = 
\begin{cases} 0 & m=0 \\ 
-\frac{s^2}{m^2}\lambda_m^{L_x}\lambda_s^{L_x}\int_0^{L_x}{\rm sin}\bigl(\lambda_s^{L_x}x\bigr){\rm cos}\bigl(\lambda_m^{L_x}x\bigr){\rm d}x & m \neq 0 
\end{cases},
\end{aligned}
\end{equation}
where the subscripts $m,s=0,1,2,\cdots,M$. Based on Eqs. (\ref{eq:Gammaxy}), (\ref{eq:Integral1}) and (\ref{eq:Integral2}), the derivation gives the following formulas,

\renewcommand{\theequation}{\arabic{equation}a, b}
\begin{equation}\label{eq:Gammaxyxy}
\int_0^{L_x}\int_0^{L_y}{\mathbf\Gamma}_{xy}{\mathbf\Gamma}_{xy}^T{\rm d}y{\rm d}x = {\mathbf A} \ast {\mathbf B} \ast \int_0^{L_x}\int_0^{L_y}{\mathbf\Gamma}{\mathbf\Gamma}^T{\rm d}y{\rm d}x \Rightarrow {\mathbf\Gamma}_{xy}{\mathbf\Gamma}_{xy}^T = {\mathbf A} \ast {\mathbf B} \ast \bigl({\mathbf\Gamma}{\mathbf\Gamma}^T\bigr),
\end{equation}
where the symmetric matrices $\mathbf A$ and $\mathbf B$ both have the same dimensions as ${\bm\Gamma}{\bm\Gamma}^T$ does. Besides, they are independent of the spatial variables $x$ or $y$. Using the four constant vectors (i.e., $\mathbf{q}_M^{}$, $\mathbf{q}_N^{}$, $\mathbf{1}_M^{}$, and $\mathbf{1}_N^{}$), we obtain the coefficient matrices $\mathbf A$ and $\mathbf B$ given in \ref{sec:B1}.

Consequently, we have the relations between the matrix ${\bm\Gamma}{\bm\Gamma}^T$ and the five matrices, i.e., $\bm\Gamma_{xx}\bm\Gamma_{xx}^T$, $\bm\Gamma_{yy}\bm\Gamma_{yy}^T$, $\bm\Gamma_{xx}\bm\Gamma_{yy}^T$, $\bm\Gamma_{yy}\bm\Gamma_{xx}^T$, and $\bm\Gamma_{xy}\bm\Gamma_{xy}^T$. To finally calculate the stiffness and mass matrices, Three integrals given in \ref{sec:C1} are used for the calculation of $\iint_{S_0}\bm\Gamma\bm\Gamma^T{\rm d}S$, $\iint_{S_0}\bm\Gamma_x\bm\Gamma_x^T{\rm d}S$, and $\iint_{S_0}\bm\Gamma_y\bm\Gamma_y^T{\rm d}S$, where $S_0$ indicates the area of the plate.

%%---------------------------%%%%%%%%%%%%%%%%Subsection 3.2%%%%%%%%%%%%%%%
\subsection{Calculation of radiation impedance matrices}
\label{subsec:CRIM}

The computations of Eqs. (\ref{eq:RadiationMatrix_baffled}), (\ref{eq:Matrix_Mhat}) and (\ref{eq:Matrix_Khat}) need to calculate the quadruple integral. Because the proposed formulation incorporates only sine and cosine functions, the quadruple integrals can be conveniently reduced to double integrals by using variable substitution method \cite{Nelisse1998, Rhazi2010},

\renewcommand{\theequation}{\arabic{equation}}
\begin{equation}\label{eq:IntegralReducedGamma} 
\iint\limits_{S_0}\iint\limits_{S_0}\Bigl[{\mit\Gamma}_{i,mn}(\mathbf{r})G(\mathbf{r};\mathbf{r'}){\mit\Gamma}_{j,pq}^T(\mathbf{r'})\Bigr]{\rm d}\mathbf{r}{\rm d}\mathbf{r'} = \frac{L_x L_y^2}{2\pi}\int_{0}^{1}\int_{0}^{1}I_{ij,mp}(\alpha) \kappa(\alpha,\alpha') J_{ij,nq}(\alpha'){\rm d}\alpha{\rm d}\alpha',
\end{equation}

\begin{equation}\label{eq:IntegralReducedLambda} 
\iint\limits_{S_0}\iint\limits_{S_0}\Bigl[{\mit\Lambda}_{mn}(\mathbf{r})G(\mathbf{r};\mathbf{r'}){\mit\Lambda}_{pq}^T(\mathbf{r'})\Bigr]{\rm d}\mathbf{r}{\rm d}\mathbf{r'} = \frac{L_x L_y^2}{2\pi}\int_{0}^{1}\int_{0}^{1}I_{mp}^{\mit\Lambda}(\alpha) \kappa(\alpha,\alpha') J_{nq}^{\mit\Lambda}(\alpha'){\rm d}\alpha{\rm d}\alpha',
\end{equation}

\begin{equation}\label{eq:IntegralReducedLambdax} 
\iint\limits_{S_0}\iint\limits_{S_0}\Bigl[{\mit\Lambda}_{x,mn}(\mathbf{r})G(\mathbf{r};\mathbf{r'}){\mit\Lambda}_{x',pq}^T(\mathbf{r'})\Bigr]{\rm d}\mathbf{r}{\rm d}\mathbf{r'} = \frac{mp\pi L_y^2}{2L_x}\int_{0}^{1}\int_{0}^{1}I_{mp}^{\mit\Lambda_x}(\alpha) \kappa(\alpha,\alpha') J_{nq}^{\mit\Lambda}(\alpha'){\rm d}\alpha{\rm d}\alpha',
\end{equation}

\begin{equation}\label{eq:IntegralReducedLambday} 
\iint\limits_{S_0}\iint\limits_{S_0}\Bigl[{\mit\Lambda}_{y,mn}(\mathbf{r})G(\mathbf{r};\mathbf{r'}){\mit\Lambda}_{y',pq}^T(\mathbf{r'})\Bigr]{\rm d}\mathbf{r}{\rm d}\mathbf{r'} = \frac{nq\pi L_x}{2}\int_{0}^{1}\int_{0}^{1}I_{mp}^{\mit\Lambda}(\alpha) \kappa(\alpha,\alpha') J_{nq}^{\mit\Lambda_y}(\alpha'){\rm d}\alpha{\rm d}\alpha',
\end{equation}
with

\renewcommand{\theequation}{\arabic{equation}a, b}
\begin{equation}\label{eq:TildeR} 
\kappa(\alpha,\alpha') = {{\rm exp}(-{\rm j}k_0 L_x \tilde{R})}/{\tilde{R}},\quad \tilde{R} = \sqrt{\alpha^2+(L_y/L_x)^2\alpha'^2},
\end{equation}
where $\mathbf{r}=x\mathbf{i}+y\mathbf{j}$, $\mathbf{r'}=x'\mathbf{i}+y'\mathbf{j}$; the subscripts $i,j=1,2,3$; $m,p=0,1,2,\cdots,M$; $n,q=0,1,2,\cdots,N$. 

First, look at Eq. (\ref{eq:IntegralReducedGamma}) for the calculation of matrix $\mathbf{Z}_{\rm bf}$. Taking $i=j=1$ for example, we have the following formulations,

\begin{equation}\label{eq:I11J11} 
I_{11,mp}(\alpha) = H_{mp}^{\rm cc}(\alpha) + H_{pm}^{\rm cc}(\alpha), \quad J_{11,nq}(\alpha) = H_{nq}^{\rm cc}(\alpha) + H_{qn}^{\rm cc}(\alpha),
\end{equation}
with

\renewcommand{\theequation}{\arabic{equation}}
\begin{equation}\label{eq:Intcoscos} 
H_{mp}^{\rm cc}(\alpha) = \int_0^{1-\alpha}{\rm cos}(m\pi(\alpha+\beta)){\rm cos}(p\pi\beta){\rm d}\beta,
\end{equation}
which has a closed solution. The matrix $\mathbf{Z}_{\rm bf}$ obtained by integration is symmetric according to the left side of Eq. (\ref{eq:IntegralReducedGamma}). The symmetry is also reflected from Eq. (\ref{eq:I11J11}). Thus, only 6 of the 9 blocks (i.e., $3\times3$) need calculations. According to the admissible functions in Eq. (\ref{eq:DisSeries}), however, there are only 4 different integral formulas left to calculate, i.e., $H_{kl}^{\rm cc}$, $H_{kl}^{\rm cs}$, $H_{kl}^{\rm sc}$, $H_{kl}^{\rm ss}$ with $k,l$ being replaced by $m,p,n,q$. Table \ref{tab:correspondance} lists the correspondance between the 4 integrals and the entries of the 6 matrix blocks in Eq. (\ref{eq:IntegralReducedGamma}). The argument $\alpha$ of the functions is omitted for brevity. The 4 integrals have closed solutions given in \ref{sec:D1}.

\begin{table}[H]
\newcommand{\tabincell}[2]{\begin{tabular}{@{}#1@{}}#2\end{tabular}}
\setlength{\abovecaptionskip}{2pt}
\setlength{\belowcaptionskip}{2pt}
\centering
\caption{Correspondance between the entries ($I_{ij,mp}$, $J_{ij,nq}$) of the 6 matrix blocks in Eq. (\ref{eq:IntegralReducedGamma}) and the 4 integrals ($H_{kl}^{\rm cc}$, $H_{kl}^{\rm cs}$, $H_{kl}^{\rm sc}$, $H_{kl}^{\rm ss}$ with $k,l$ being replaced by $m,p,n,q$).}
\label{tab:correspondance}
\scalebox{1.0}{
\begin{tabular}{@{}lcccccc@{}}
\toprule[1pt]
$(i,j)$ & (1,1) & (1,2) & (1,3) & (2,2) & (2,3) & (3,3) \\
\midrule[1pt]
$I_{ij,mp}=$ & $H_{mp}^{\rm cc} + H_{pm}^{\rm cc}$ & $H_{mp}^{\rm cc} + H_{pm}^{\rm cc}$ & $H_{mp}^{\rm cs} + H_{pm}^{\rm sc}$ & $H_{mp}^{\rm cc} + H_{pm}^{\rm cc}$ & $H_{mp}^{\rm cs} + H_{pm}^{\rm sc}$ & $H_{mp}^{\rm ss} + H_{pm}^{\rm ss}$ \\
$J_{ij,nq}=$ & $H_{nq}^{\rm cc} + H_{qn}^{\rm cc}$ & $H_{nq}^{\rm cs} + H_{qn}^{\rm sc}$ & $H_{nq}^{\rm cc} + H_{qn}^{\rm cc}$ & $H_{nq}^{\rm ss} + H_{qn}^{\rm ss}$ & $H_{nq}^{\rm sc} + H_{qn}^{\rm cs}$ & $H_{nq}^{\rm cc} + H_{qn}^{\rm cc}$ \\
\bottomrule[1pt]
\end{tabular}}
\end{table}

Then, Look at Eqs. (\ref{eq:IntegralReducedLambda})-(\ref{eq:IntegralReducedLambday}) for the calculation of matrices $\hat{\mathbf M}$ and $\hat{\mathbf K}$. We have derived the following formulas, $I_{mp}^{\mit\Lambda}=I_{33,mp}$, $J_{nq}^{\mit\Lambda}=J_{22,nq}$, $I_{mp}^{\mit\Lambda_x}=I_{11,mp}$, $J_{nq}^{\mit\Lambda_y}=J_{11,nq}$, which can be calculated according to the formulations in Table \ref{tab:correspondance}.

Finally, the matrices $\mathbf{Z}_{\rm bf}$, $\hat{\mathbf M}$, and $\hat{\mathbf K}$, corresponding to Eq. (\ref{eq:RadiationMatrix_baffled}), (\ref{eq:Matrix_Mhat}), and (\ref{eq:Matrix_Khat}), respectively, can be evaluated through a Gaussian numerical integration scheme. The quadruple integration has been reduced to twofold integration.

%%---------------------------%%%%%%%%%%%%%%%%Subsection 3.3%%%%%%%%%%%%%%%
\subsection{Calculation of modal forces and responses}
\label{subsec:CMF}

\textbf{(1) Single-point force and mean square velocity}

The external work under a single-point force is obtained by integrating the product of the Dirac function and the displacement over the entire area of the plate. Assuming that a single-point force has an amplitude of $F_0$, the calculated modal force gives the following form, $\mathbf{f}=F_0\mathbf{\Gamma}(x_0,y_0)$, where $(x_0,y_0)$ is the forced location. 

After solving the forced vibration equation, $\bigl(\mathbf{K}-\omega^2\mathbf{M}\bigr)\mathbf{a}=\mathbf{f}$, Eq. (\ref{eq:MotionEq_baffled}), or Eq. (\ref{eq:MotionEq_unbaffled}), the mean square velocity $\langle v^2\rangle$ is calculated by

\begin{equation}\label{eq:MSV} 
\langle v^2\rangle = \frac{\omega^2}{L_x L_y}\iint_{S_0} |w|^2{\rm d}S = \frac{\omega^2}{L_x L_y}\bigg|\mathbf{a}^H\Bigl(\iint_{S_0} \mathbf\Gamma \mathbf\Gamma^T{\rm d}S\Bigr)\mathbf{a}\bigg|,
\end{equation}
where the superscript $H$ indicates the conjugate transpose of a vector or matrix. The mean square velocity level (MSVL) is given by $L_{v}=10{\rm lg}(\langle v^2\rangle/v_{\rm ref}^2)$ with $v_{\rm ref}$ the reference velocity.

\textbf{(2) Obliquely incident plane wave and the transmission loss}

The obliquely incident plane wave with an incident angle $\theta_{\rm in}$ and an azimuth angle $\varphi_{\rm in}$ is applied at the incident surface of the plate. The plane wave has an incident pressure as

\begin{equation}\label{eq:p_oblique} 
p_{\rm in}(x,y,z)|_{z=0} = P_{\rm in}{\rm exp}(-{\rm j}k_0 x{\rm sin}\theta_{\rm in}{\rm cos}\varphi_{\rm in}-{\rm j}k_0 y{\rm sin}\theta_{\rm in}{\rm sin}\varphi_{\rm in}),
\end{equation}
where $P_{\rm in}$ is the incident pressure amplitude and a harmonic time dependence of the form ${\rm exp}({\rm j}\omega t)$ is assumed. We may first write the expression of the external work done by the blocked pressure of the obliquely incident plane wave as

\begin{equation}\label{eq:Wext_oblique} 
W_{\rm ext} = \iint_{S_0} p_{\rm bl}^{}(x,y)w(x,y){\rm d}S = \biggl[\iint_{S_0}2p_{\rm in}^{}(x,y,0)\mathbf\Gamma(x,y){\rm d}S\biggr]^T\mathbf{a} = \mathbf{f}_{\rm bl}^T\mathbf{a},
\end{equation}
where $\mathbf{f}_{\rm bl}$ is the modal force of the block pressure $p_{\rm bl}$, which is twice the incident pressure. There are corresponding blocks for the modal force $\mathbf{f}_{\rm bl}^{}=[\mathbf{f}_{\rm bl1}^T,\mathbf{f}_{\rm bl2}^T,\mathbf{f}_{\rm bl3}^T]^T$ according to the blocks of $\mathbf\Gamma$. After the integration operation, we get the entries as follows:

\renewcommand{\theequation}{\arabic{equation}a, b, c}
\begin{gather}\label{eq:fbl1-3}
f_{{\rm bl1},mn} = 2|P_{\rm in}| Y_{L_x,m}^{\rm c}Y_{L_y,n}^{\rm c}, \quad f_{{\rm bl2},mn} = 2|P_{\rm in}| Y_{L_x,m}^{\rm c}Y_{L_y,n}^{\rm s}, \quad f_{{\rm bl3},mn} = 2|P_{\rm in}| Y_{L_x,m}^{\rm s}Y_{L_y,n}^{\rm c},
\end{gather}
where the four terms $ Y_{L_x,m}^{\rm c}$, $Y_{L_y,n}^{\rm c}$, $Y_{L_x,m}^{\rm s}$, $Y_{L_y,n}^{\rm s}$ are computed through four special integrations as given in \ref{sec:E1}.

By solving the forced vibration equation, e.g., Eq. (\ref{eq:MotionEq_baffled}), with $\mathbf{f}=\mathbf{f}_{\rm bl}^{}$, the complex amplitudes of the generalized displacement $\mathbf{a}$ can be obtained for a specific obliquely incident plane wave. The radiation impedance matrix is calculated by a reduced-order integration. The radiated sound power is then obtained by 

\renewcommand{\theequation}{\arabic{equation}}
\begin{equation}\label{eq:RadiatedPowerOblique}
{\mit\Pi}_{\rm rad}(\theta_{\rm in},\varphi_{\rm in}) = \frac{\omega^2}{2}{\rm Re}(\mathbf{a}^H\mathbf{Z}_{\rm bf}\mathbf{a}).
\end{equation}
Utilizing the incident sound power of an obliquely plane wave, ${\mit\Pi}_{\rm in}=|P_{\rm in}|^2S_0 {\rm cos}\theta_{\rm in}/(2\rho_0 c_0)$, the sound transmission coefficient $\tau$ of the plate follows from Eq. (\ref{eq:RadiatedPowerOblique}) as 

\begin{equation}\label{eq:TauOblique}
\tau = \frac{{\mit\Pi}_{\rm rad}}{{\mit\Pi}_{\rm in}} = \frac{\rho_0 c_0\omega^2}{|P_{\rm in}|^2S_0 {\rm cos}\theta_{\rm in}}{\rm Re}(\mathbf{a}^H\mathbf{Z}_{\rm bf}\mathbf{a}).
\end{equation}
Then, the sound TL of the plate is given by ${TL}=-10{\rm lg}\tau$.

\textbf{(3) Transmission loss under diffuse acoustic field}

The radiated sound power under excitation of arbitrary oblique incoming plane wave can be obtained by solving Eq. (\ref{eq:RadiatedPowerOblique}). Integrating this radiated power in hemispherical space gives the radiated sound power under DAF excitation, viz.,

\begin{equation}\label{eq:RadiatedPower_DAF}
{\mit\Pi}_{\rm rad, DAF} = \int_{0}^{\theta_{\rm lim}}\int_{0}^{2\pi}{\mit\Pi}_{\rm rad}(\theta_{\rm in},\varphi_{\rm in})\sin\theta_{\rm in}{\rm d}\varphi_{\rm in}{\rm d}\theta_{\rm in},
\end{equation}
where $\theta_{\rm lim}$ indicates a limited incident angle of DAF, here taken as $\pi/2$. Gaussian integration is used to solve the twofold integral. The incident sound power of the DAF is given by

\begin{equation}\label{eq:IncidentPower_DAF}
{\mit\Pi}_{\rm in, DAF} = \int_{0}^{\pi/2}\int_{0}^{2\pi}{\mit\Pi}_{\rm in}\sin\theta_{\rm in}{\rm d}\varphi_{\rm in}{\rm d}\theta_{\rm in} = |P_{\rm in}|^2S_0 \pi /(2\rho_0 c_0).
\end{equation}
According Eq. (\ref{eq:RadiatedPower_DAF}) and (\ref{eq:IncidentPower_DAF}), the sound TL is obtained as follows,

\begin{equation}\label{eq:TL_DAF}
{TL}_{\rm DAF} = -10{\rm lg}\Bigl(\frac{{\mit\Pi}_{\rm rad, DAF}}{{\mit\Pi}_{\rm in, DAF}}\Bigr) = -10{\rm lg}\Bigl(\frac{2\rho_0 c_0}{|P_{\rm in}|^2S_0 \pi}{\mit\Pi}_{\rm rad, DAF}\Bigr).
\end{equation}

%%---------------------------%%%%%%%%%%%%%%%Section 4%%%%%%%%%%%%%%%%%%---------------------------------
\section{Results and discussion}
\label{sec:Results} 

The natural frequencies of 21 combinations of classical boundary conditions are calculated by the proposed method and compared to the literature results (Section \ref{subsec:NF}), followed by vibrating response of an unbaffled plate under single-point force in Section \ref{subsec:MSV_unbaffled}. The sound TLs of a baffled plate with different boundary conditions under the plane wave excitation and DAF are compared with FE-BEM simulations (Section \ref{subsec:TL_baffled_oblique}) and FE-SEA simulations (Section \ref{subsec:TL_baffled_DAF}), respectively.

%%---------------------------%%%%%%%%%%%%%%%%Subsection 4.1%%%%%%%%%%%%%%%
\subsection{Natrural frequencies and vibrating modes}
\label{subsec:NF}

Classical boundary conditions are regarded with free (F), simply-supported (S), and clamped (C) edges. The symbol combination represents the boundary conditions of edges in clockwise order, e.g., C-F-S-F standing for clamped, free, simply-supported, and free edges of a rectangular plate. Both transverse and rotational spring constants of a free edge are zero. The transverse spring constant of a simply-supported edge takes a very large number while the rotational spring constant is zero. A clamped edge can be seen as a special case when both the transverse and rotational spring constants are infinitely large \cite{Li2009}. We use two very large numbers, i.e., $D/(L_x L_y)^{3/2}\times 10^6$ and $D/(L_x L_y)^{1/2}\times 10^6$ for the transverse and rotational spring constants, respectively. Poisson's ratio of the plate takes 0.3 for the calculation because the dimensionless frequency $\tilde{\omega}=\omega L_x^2\sqrt{\rho h/D}$ is dependent on Poisson's ratio other than bulk density, Young's modulus, and thickness. 

The first six frequencies are calculated to compare with the documented results in the literature involving 21 combinations of the traditional boundary conditions. To examine the convergence of solution, we compare the first six dimensionless frequencies $\tilde{\omega}$ of a square plate ($r=L_x/L_y=1$) in Table \ref{tab:convergence} using different numbers of terms in the series expansions for the bending displacement. It is shown that all the six frequencies converge at $M=N=9$ for a given five-digit precision. Then, the first six dimensionless frequencies of a clamped plate with 5 different aspect ratios ($r$ = 1.0, 1.5, 2.0, 2.5 and 3.0) are predicted by setting $M=N=10$ (Table \ref{tab:CCCC}). The predicted results display a good agreement with those previously reported in Refs. \cite{Leissa1973, Blevins1979, Li2009}.

\begin{table}[H]
\newcommand{\tabincell}[2]{\begin{tabular}{@{}#1@{}}#2\end{tabular}}
\setlength{\abovecaptionskip}{2pt}
\setlength{\belowcaptionskip}{2pt}
\centering
\caption{Dimensionless frequencies $\tilde{\omega}$ for a C-C-C-C square plate ($r=L_x/L_y=1$) with the increase of truncating numbers ($M=N$).}
\label{tab:convergence}
\scalebox{0.9}{
\begin{tabular}{@{}lcccccc@{}}
\toprule[2pt]
\multirow{2}{*}{$M=N$} & \multicolumn{6}{c}{$\tilde{\omega}=\omega L_x^2\sqrt{\rho h/D}$} \\
\cline{2-7}
& 1 & 2 & 3 & 4 & 5 & 6 \\
\midrule[2pt]
6 & 35.982 & 73.374 & 73.374 & 108.36 & 131.50 & 132.14 \\
7 & 35.979 & 73.371 & 73.387 & 108.15 & 131.50 & 132.13 \\
8 & 35.979 & 73.368 & 73.368 & 108.15 & 131.50 & 132.13 \\
9 & 35.979 & 73.367 & 73.367 & 108.15 & 131.50 & 132.13 \\
10 & 35.979 & 73.367 & 73.367 & 108.15 & 131.50 & 132.13 \\
11 & 35.979 & 73.367 & 73.367 & 108.15 & 131.50 & 132.13 \\
12 & 35.979 & 73.367 & 73.367 & 108.15 & 131.50 & 132.13 \\
\bottomrule[2pt]
\end{tabular}}
\end{table}

\begin{table}[H]
\newcommand{\tabincell}[2]{\begin{tabular}{@{}#1@{}}#2\end{tabular}}
\setlength{\abovecaptionskip}{2pt}
\setlength{\belowcaptionskip}{2pt}
\centering
\caption{Dimensionless frequencies $\tilde{\omega}$ for a C-C-C-C plate in 5 different aspect ratios with a truncating number of the terms $M=N=10$.}
\label{tab:CCCC}
\scalebox{0.9}{
\begin{tabular}{@{}lcccccc@{}}
\toprule[2pt]
\multirow{2}{*}{$r=L_x/L_y$} & \multicolumn{6}{c}{$\tilde{\omega}=\omega L_x^2\sqrt{\rho h/D}$} \\
\cline{2-7}
& 1 & 2 & 3 & 4 & 5 & 6 \\
\midrule[2pt]
\multirow{4}{*}{1.0} & 35.98 & 73.39 & 73.39 & 108.2 & 131.6 & 132.2 \\
& 35.99 \cite{Blevins1979} & 73.41 & 73.41 & 108.3 & 131.6 & 132.2 \\
& 35.99 \cite{Leissa1973} & 73.39 & 73.39 & 108.2 & 131.6 & 132.2 \\
& 35.99 \cite{Li2009} & 73.39 & 73.39 & 108.2 & 131.6 & 132.2 \\
\cline{2-7}
\multirow{2}{*}{1.5} & 60.76 & 94.03 & 148.8 & 149.7 & 179.6 & 226.8 \\
& 60.76 \cite{Li2009} & 93.83 & 148.8 & 149.7 & 179.6 & 226.8 \\
\cline{2-7}
\multirow{2}{*}{2.0} & 98.32 & 127.3 & 179.1 & 253.3 & 255.9 & 284.3 \\
& 98.31 \cite{Li2009} & 127.3 & 179.1 & 253.3 & 255.9 & 284.3 \\
\cline{2-7}
\multirow{3}{*}{2.5} & 147.8 & 173.8 & 221.4 & 291.7 & 384.3 & 394.3 \\
& 147.8 \cite{Blevins1979} & 173.9 & 221.5 & 291.9 & 384.7 & 394.4 \\
& 147.8 \cite{Li2009} & 173.8 & 221.4 & 291.7 & 384.3 & 394.2 \\
\cline{2-7}
\multirow{2}{*}{3.0} & 208.7 & 232.7 & 276.7 & 342.9 & 431.7 & 542.8 \\
& 208.8 \cite{Li2009} & 232.7 & 276.7 & 342.8 & 431.6 & 542.7 \\
\bottomrule[2pt]
\end{tabular}}
\end{table}

\begin{table}[H]
\newcommand{\tabincell}[2]{\begin{tabular}{@{}#1@{}}#2\end{tabular}}
\setlength{\abovecaptionskip}{2pt}
\setlength{\belowcaptionskip}{2pt}
\centering
\caption{Dimensionless frequencies $\tilde{\omega}$ for a square plate ($r=1$) under 20 different boundary conditions with a truncating number of terms $M=N=10$.}
\label{tab:20examples}
\scalebox{0.9}{
\begin{tabular}{@{}lcccccc@{}}
\toprule[2pt]
\multirow{2}{*}{Boundary types} & \multicolumn{6}{c}{$\tilde{\omega}=\omega L_x^2\sqrt{\rho h/D}$} \\
\cline{2-7}
& 1 & 2 & 3 & 4 & 5 & 6 \\
\midrule[2pt]
\multirow{2}{*}{F-F-F-F} & 13.46 & 19.60 & 24.27 & 34.80 & 34.80 & 61.09 \\
& 13.49 \cite{Blevins1979} & 19.79 & 24.43 & 35.02 & 35.02 & 61.53 \\
\cline{2-7}
\multirow{2}{*}{S-F-F-F} & 6.649 & 14.90 & 25.39 & 26.00 & 48.48 & 50.58 \\
& 6.648 & 15.02 & 25.49 & 26.13 & 48.71 & 50.85 \\
\cline{2-7}
\multirow{2}{*}{C-F-F-F} & 3.471 & 8.505 & 21.28 & 27.20 & 30.95 & 54.17 \\
& 3.492 & 8.525 & 21.43 & 27.33 & 31.11 & 54.44 \\
\cline{2-7}
\multirow{2}{*}{S-F-F-S} & 3.372 & 17.33 & 19.30 & 38.25 & 51.03 & 53.52 \\
& 3.369 & 17.41 & 19.37 & 38.29 & 51.32 & 53.74 \\
\cline{2-7}
\multirow{2}{*}{S-F-S-F} & 9.633 & 16.15 & 36.77 & 38.95 & 46.78 & 70.83 \\
& 9.631 & 16.14 & 36.73 & 38.95 & 46.74 & 70.74 \\
\cline{2-7}
\multirow{2}{*}{C-F-S-F} & 15.19 & 20.59 & 39.75 & 49.44 & 56.28 & 77.36 \\
& 15.29 & 20.67 & 39.78 & 49.73 & 56.62 & 77.37 \\
\cline{2-7}
\multirow{2}{*}{C-F-F-S} & 5.352 & 19.08 & 24.67 & 43.10 & 52.71 & 63.75 \\
& 5.364 & 19.17 & 24.77 & 43.19 & 53.00 & 64.05 \\
\cline{2-7}
\multirow{2}{*}{C-F-C-F} & 22.16 & 26.39 & 43.58 & 61.15 & 67.13 & 79.80 \\
& 22.27 & 26.53 & 43.66 & 61.47 & 67.55 & 79.90 \\
\cline{2-7}
\multirow{2}{*}{C-F-F-C} & 6.918 & 23.90 & 26.58 & 47.64 & 62.69 & 65.51 \\
& 6.942 & 24.03 & 26.68 & 47.78 & 63.04 & 65.83 \\
\cline{2-7}
\multirow{2}{*}{S-F-S-S} & 11.70 & 27.80 & 41.23 & 59.16 & 61.93 & 90.31 \\
& 11.68 & 27.76 & 41.20 & 59.07 & 61.86 & 90.29 \\
\cline{2-7}
\multirow{2}{*}{S-F-S-C} & 12.69 & 33.09 & 41.71 & 63.05 & 72.42 & 90.62 \\
& 12.69 & 33.07 & 41.70 & 63.01 & 72.40 & 90.61 \\
\cline{2-7}
\multirow{2}{*}{C-F-C-S} & 23.36 & 35.56 & 62.84 & 66.74 & 77.33 & 108.8 \\
& 23.46 & 35.61 & 63.13 & 66.81 & 77.50 & 109.0 \\
\cline{2-7}
\multirow{2}{*}{C-F-S-S} & 16.80 & 31.14 & 51.39 & 64.06 & 67.57 & 101.2 \\
& 16.87 & 31.14 & 51.63 & 64.04 & 67.65 & 101.2 \\
\cline{2-7}
\multirow{2}{*}{C-F-S-C} & 17.54 & 36.03 & 51.80 & 71.07 & 74.34 & 105.8 \\
& 17.62 & 36.05 & 52.07 & 71.19 & 74.35 & 106.3 \\
\cline{2-7}
\multirow{2}{*}{C-F-C-C} & 23.91 & 39.98 & 63.19 & 76.69 & 80.53 & 116.6 \\
& 24.02 & 40.04 & 63.49 & 76.76 & 80.71 & 116.8 \\
\cline{2-7}
\multirow{2}{*}{S-S-S-S} & 19.78 & 49.42 & 49.42 & 79.18 & 98.67 & 98.84 \\
& 19.74 & 49.35 & 49.35 & 78.96 & 98.70 & 98.70 \\
\cline{2-7}
\multirow{2}{*}{S-S-S-C} & 23.67 & 51.72 & 58.67 & 86.24 & 100.3 & 113.2 \\
& 23.65 & 51.67 & 58.65 & 86.13 & 100.3 & 113.2 \\
\cline{2-7}
\multirow{2}{*}{C-S-S-C} & 27.07 & 60.52 & 60.83 & 92.87 & 114.5 & 114.7 \\
& 27.06 & 60.54 & 60.79 & 92.86 & 114.6 & 114.7 \\
\cline{2-7}
\multirow{2}{*}{S-C-S-C} & 28.95 & 54.73 & 69.30 & 94.54 & 102.2 & 129.0 \\
& 28.95 & 54.74 & 69.32 & 94.59 & 102.2 & 129.1 \\
\cline{2-7}
\multirow{2}{*}{C-S-C-C} & 31.82 & 63.31 & 71.05 & 100.7 & 116.3 & 130.3 \\
& 31.83 & 63.35 & 71.08 & 100.8 & 116.4 & 130.4 \\
\bottomrule[2pt]
\end{tabular}}
\end{table}

In addition, 20 more combined boundary conditions for the four edges of a rectangular plate ($r=1$) are listed in Table \ref{tab:20examples}. The predicted dimensionless frequencies $\tilde{\omega}$ using our method are shown in the first line for each case. The results reported by Blevins \cite{Blevins1979} are shown in the second line. Good agreements are achieved. 

\begin{figure}
\centering
\includegraphics[scale=0.13]{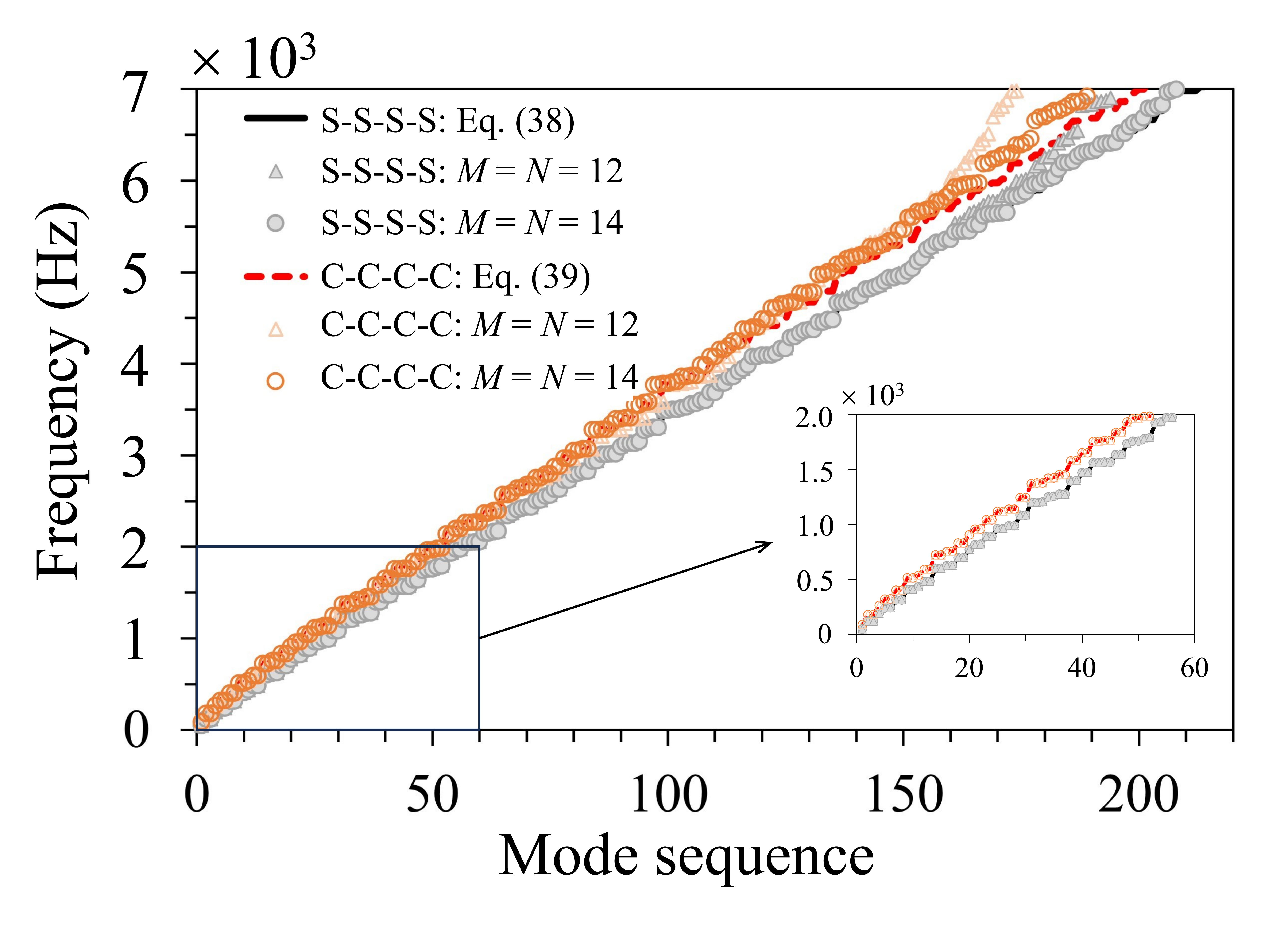}
\caption{A comparison with exact solutions for the first 200 natural frequencies.}
\label{fig:Mode_Sequency}
\end{figure}

The first 200 natural frequencies predicted by the proposed method are compared with the exact solutions for S-S-S-S and C-C-C-C square plates. The thin rectangular plate has dimensions of $L_x = L_y = 1\ \rm m$, density of $\rho=7800\ \rm kg\ m^{-3}$, thickness of $h = 0.01\ \rm m$, Poisson's ratio of $\mu=0.3$, and Young's modulus of $E_0=2\times10^{11}\ \rm Pa$. The formulas to calculate the frequencies corresponding to the simply-supported and clamped edges are given as \cite{Blevins1979},

\begin{equation}\label{eq:SSSS}
f_{mn}^{\rm SSSS} = \frac{\pi}{2} \sqrt{\frac{D}{\rho h}} \Biggl[\Bigl(\frac{m}{L_x}\Bigr)^2+\Bigl(\frac{n}{L_y}\Bigr)^2\Biggr],
\end{equation}

\begin{equation}\label{eq:CCCC}
f_{mn}^{\rm CCCC} = \frac{\pi}{2} \sqrt{\frac{D}{\rho h}} \sqrt{\Bigl(\frac{G_x}{L_x}\Bigr)^4+\Bigl(\frac{G_y}{L_y}\Bigr)^4+\frac{2 H_x H_y}{L_x^2 L_y^2}},
\end{equation}
where $m,n \ (=1,2,\cdots)$ indicate the mode index. For the C-C-C-C plate, $G_x = 1.506$ and $H_x = 1.248$, when $m = 1$; otherwise, $G_x = m+1/2$, $H_x = G_x^2[1-2/(G_x \pi)]$; $G_y$ and $H_y$ are calculated in a similar way. It can be seen from Fig. \ref{fig:Mode_Sequency} that the predicted results by the proposed method agree well with the exact solutions within the first 100 modes. For the higher order modes, noticeable deviations are observed, and the predictions are larger than the exact solutions because the assumed functions introduce constraints to the system according to Rayleigh principle. The relative errors between the prediction and the exact solution are less than 3\% for those modes below 3 kHz, when $M = N = 12$. Increased $M$ and $N$ (e.g., $=14$) lead to a smaller deviation for high order modes. Previous studies using wavelet-based analytical model for predictions showed a similar phenomenon \cite{Zhang2017b, Ma2018}.

\begin{figure}
\centering
\includegraphics[scale=0.096]{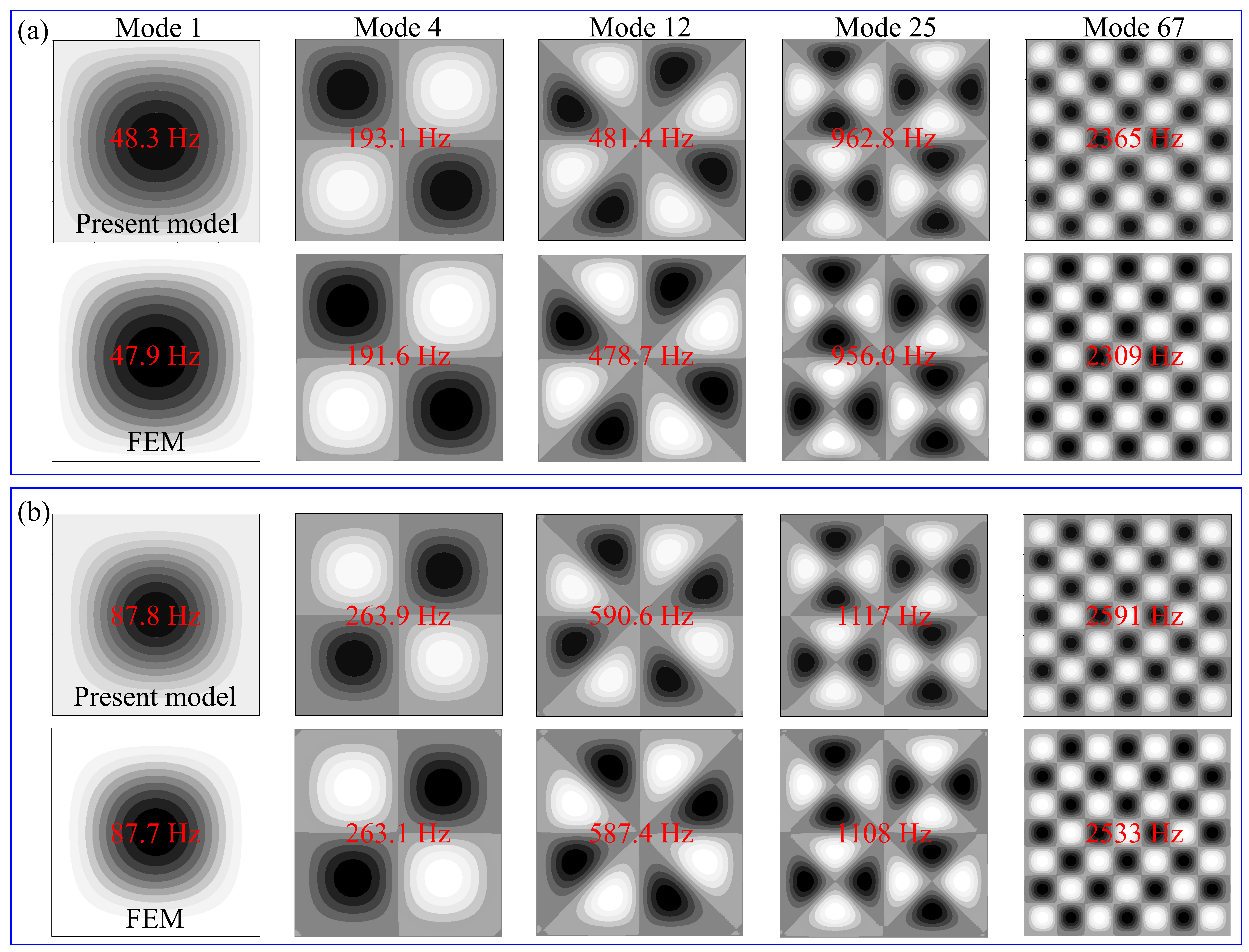}
\caption{Mode shape comparisons: (a) S-S-S-S; (b) C-C-C-C.}
\label{fig:Mode_Shapes}
\end{figure}

Predicted mode shapes of the same square plate using $M = N = 14$ are compared with the FEM simulations. The FEM model built by HyperMesh software discretizes the plate into 0.1 million quadrilateral first-order elements with an element size of 10 mm. The built-in module, OptiStruct, is used to solve the mode shapes. Consistent results are observed between the present model and the FEM simulation, as shown in Fig. \ref{fig:Mode_Shapes}. The boundary conditions (simply supported or clamped) have a greater impact on low-frequency modes (see, modes 1, 4, and 12) than the high-frequency modes (e.g., modes 25 and 67). The transverse displacement of clamped plate is concentrated in the center, while that of the simply supported plate extends to the edges due the unconstrained rotational freedom. The great effect of boundary conditions on low-frequency modes is related to their effect on low-frequency sound insulation, seen in Section \ref{subsec:TL_baffled_oblique} and \ref{subsec:TL_baffled_DAF}.

%%---------------------------%%%%%%%%%%%%%%%%Subsection 4.2%%%%%%%%%%%%%%%
\subsection{Mean square velocity of an unbaffled plate}
\label{subsec:MSV_unbaffled}

\begin{figure}
\centering
\includegraphics[scale=0.2]{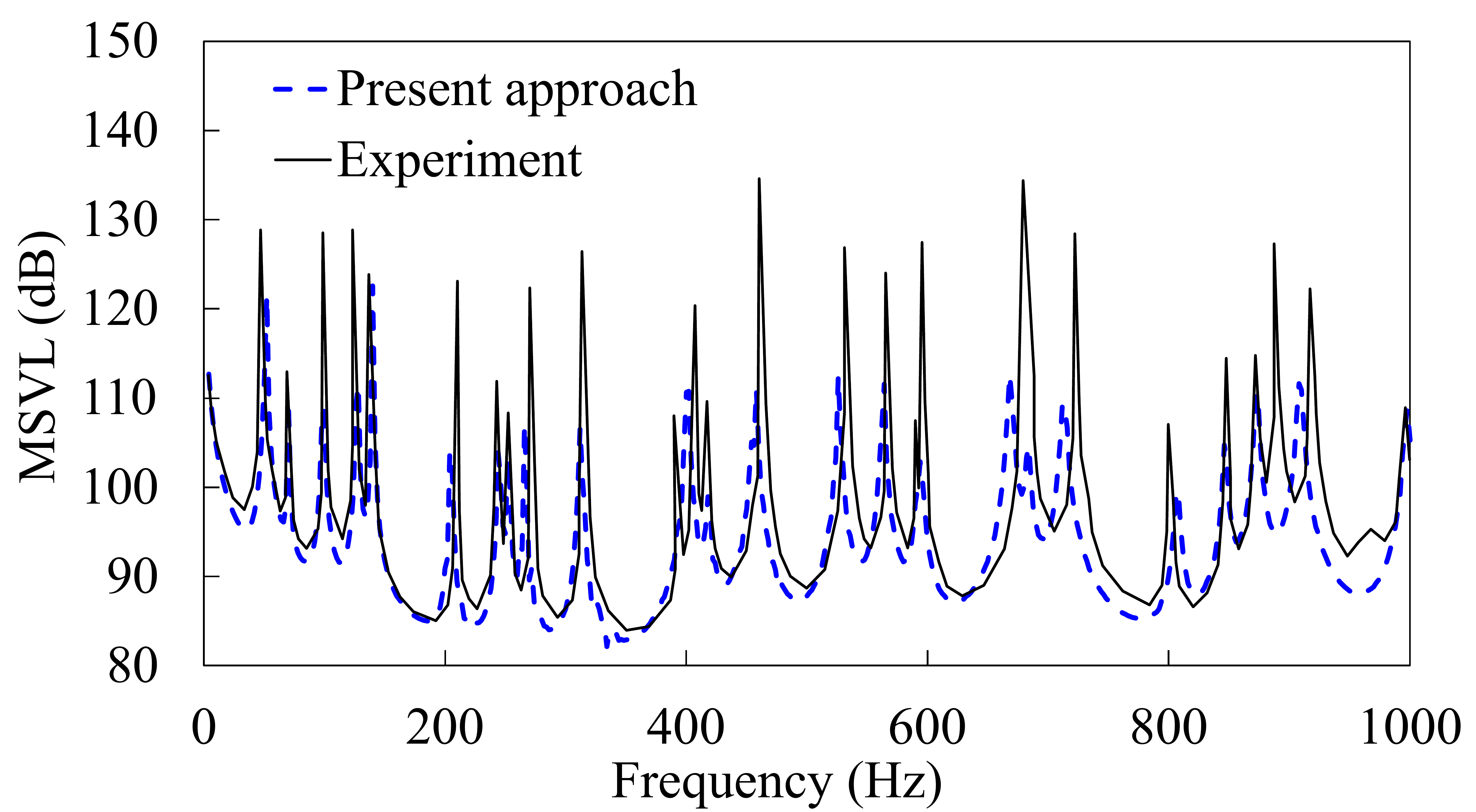}
\caption{The mean square velocity level (MSVL) of a free plate: a validation of the present approach by experimental results \cite{Nelisse1998}.}
\label{fig:MSV_comparison}
\end{figure}

The proposed method is used to predict the mean square velocity of an unbaffled plate under a single-point force. The calculations are compared with experimental results \cite{Nelisse1998}. In the experiment, a free plate with sizes of $L_x=0.48\ \rm m$, $L_y=0.42\ \rm m$ and thickness of $h=0.00322\ \rm m$ was excited by a shaker at the position $(x_0,y_0)=(0.08, 0.07)$ m. A laser vibrometer was used to measure and compute the quadratic velocity through 361 points across the plate, which has a density, Young's modulus, Poisson's ratio, and damping loss factor of $\rho=2680\ \rm kg\ m^{-3}$, $E_0=6.7\times10^{10}\ \rm Pa$, $\mu=0.3$, and $\eta=0.003$, respectively. The test was carried out in the air medium, whose damping is negligibly small. The unbaffled plate possesses free edges. In the numerical calculation, Eq. (\ref{eq:MotionEq_unbaffled}) is deployed to compute the response with air density $\rho_0=1.21\ \rm kg\ m^{-3}$, sound speed $c_0=343\ \rm m\ s^{-1}$, and zero damping. The truncated terms $M=N=8$ are enough for prediction to converge in the range below 1 kHz. Calculations of the modal force and MSVL are referred to Section \ref{subsec:CMF} with a reference velocity of $v_{\rm ref}=1\times10^{-9}\ \rm m\ s^{-1}$. Both the predicted and measured MSVL are shown in Fig. \ref{fig:MSV_comparison}, displaying good agreements between the present approach and the experimental measurement.

%%---------------------------%%%%%%%%%%%%%%%%Subsection 4.3%%%%%%%%%%%%%%%
\subsection{Transmission loss of a baffled plate under obliquely incident plane wave}
\label{subsec:TL_baffled_oblique}

The present method can be used to compute the sound TL of a rectangular plate under acoustic excitation. A selected plate has the following parameters: dimensions $L_x=$ 0.35 m, $L_y=$ 0.22 m, thickness $h=0.001$ m, density $\rho=2814 {\rm\ kg\ m^{-3}}$, Young's modulus $E_0=7.1\times10^{10}$ Pa, Poisson's ratio $\mu=0.33$, damping loss factor $\eta=0.001$. The properties of the air medium are the same as mentioned above. The acoustic excitation is the obliquely incident plane wave with an incident angle and azimuth angle of $45^\circ$ and $0^\circ$, respectively. Equation (\ref{eq:TauOblique}) is used for the semi-analytical computation with truncated terms of $M=10,\ N=9$, which were determined so that the prediction was converged below 2 kHz. The twofold integrals in Eq. (\ref{eq:IntegralReducedGamma}) are evaluated using a Gaussian numerical integration scheme in the integral space with 110 sampling points. Frequencies in 10-2000 Hz with a constant band of 10 Hz are considered. The pressure amplitude of the plane wave takes $\sqrt{2}$ Pa in the semi-analytical calculation.

The calculated TLs of the plate with three different boundary conditions, i.e., F-F-F-F, S-S-S-S, and C-C-C-C edges, are compared with FE-BEM simulations. In the simulation setup, the plate is discretized with 0.01 m quadrilateral first-order grids, producing about 1000 nodes. The incident and radiating side of the plate is each connected to a BEM fluid, which models the semi-infinite free sound field under a baffled condition. The obliquely incident plane wave with unit pressure of 1 Pa is applied to the BEM fluid of the incidence.

\begin{figure}
\centering
\includegraphics[scale=0.090]{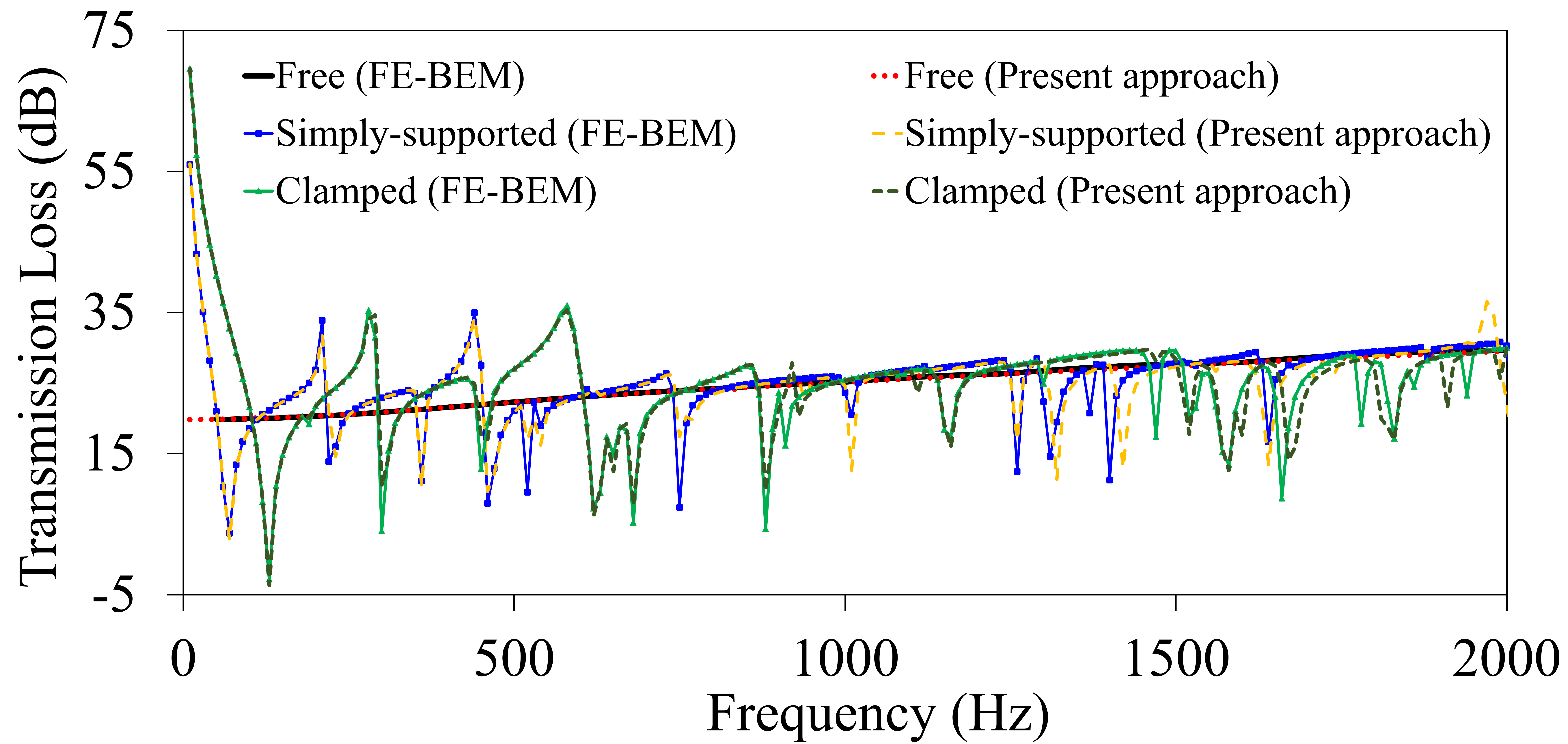}
\caption{The sound TL of a baffled plate with free, simply-supported and clamped edges under the excitation of obliquely incident plane wave: a validation of the present approach by the hybrid FE-BEM simulation.}
\label{fig:TL_oblique_comparison}
\end{figure}

A comparison of sound TLs between the present approach and the FE-BEM simulations is shown in Fig \ref{fig:TL_oblique_comparison}. Consistency of the predicted TLs is observed, demonstrating that the proposed method may successfully compute sound radiation and TL with acceptable accuracy. In the much high-frequency bands (close to 2 kHz), there exist some discrepancies due to numerical errors. Moreover, comparing the sound TLs of the three boundary conditions, we find that the boundary conditions affect the results more notably in low-frequency bands, specifically below 800 Hz. In the high-frequency bands (e.g., above 800 Hz), except for many troughs of the TL curves corresponding to the simply-supported and clamped plates, the maximum envelopes are almost the same as the TL values corresponding to the free plate. The effect of boundary conditions on the acoustic response at high frequencies is negligibly small due to the large density of high-frequency modes. They are little affected by boundary conditions, which is evident from the mode shapes given in Fig. \ref{fig:Mode_Shapes}.

The proposed method can implement vibration and sound radiation analysis of a rectangular plate with a different elastic boundary condition by merely adjusting the transverse and rotational spring constants. But the FE-BEM is difficult to achieve other elastic boundary conditions, e.g., the spring constant is neither equal to zero nor infinity, let alone spring constants with complex values.

%%---------------------------%%%%%%%%%%%%%%%%Subsection 4.4%%%%%%%%%%%%%%%
\subsection{Transmission loss of a baffled plate under DAF excitation}
\label{subsec:TL_baffled_DAF}

\begin{figure}
\centering
\includegraphics[scale=0.090]{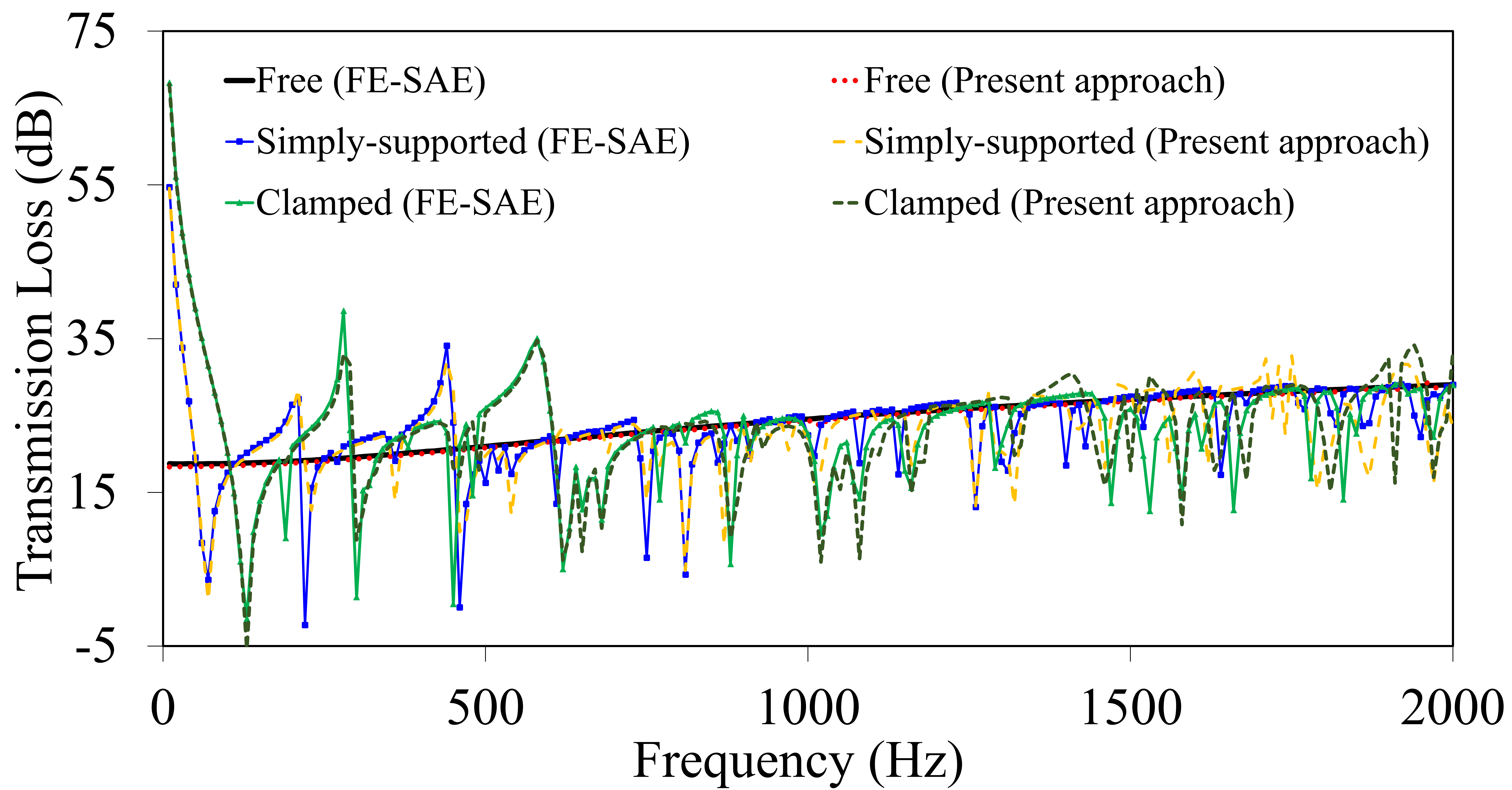}
\caption{The sound TL of a baffled plate with free, simply-supported and clamped edges under the DAF excitation: a validation of the present approach by the hybrid FE-SEA simulation.}
\label{fig:TL_DAF_comparison}
\end{figure}

A rectangular plate with the same dimensions and material parameters is used to calculate the sound TL under a DAF excitation. The semi-analytical calculation is performed by using Eq. (\ref{eq:TL_DAF}) with truncated terms of $M=10,\ N=9$, and 42 sampling points for Gaussian integration. The "VA One" software is employed to carry out hybrid FE-SEA simulations for a comparison. The plate is discretized into quadrilateral elements with about 700 nodes, whilst the source and receiver sides of the plate are modeled as SEA fluids. A DAF excitation with pressure of 1 Pa is applied to the source side. Hybrid area junctions are utilized to model the coupling interfaces between SEA fluids and the structural FE subsystem. Details of the hybrid FE-SEA approach and its applications are referred to \cite{Langley2007,Deng2021a}. Figure \ref{fig:TL_DAF_comparison} shows good agreements between our semi-analytical solutions and the FE-SEA results, with acceptable deviations at high frequencies. 

\begin{figure}
\centering
\includegraphics[scale=0.15]{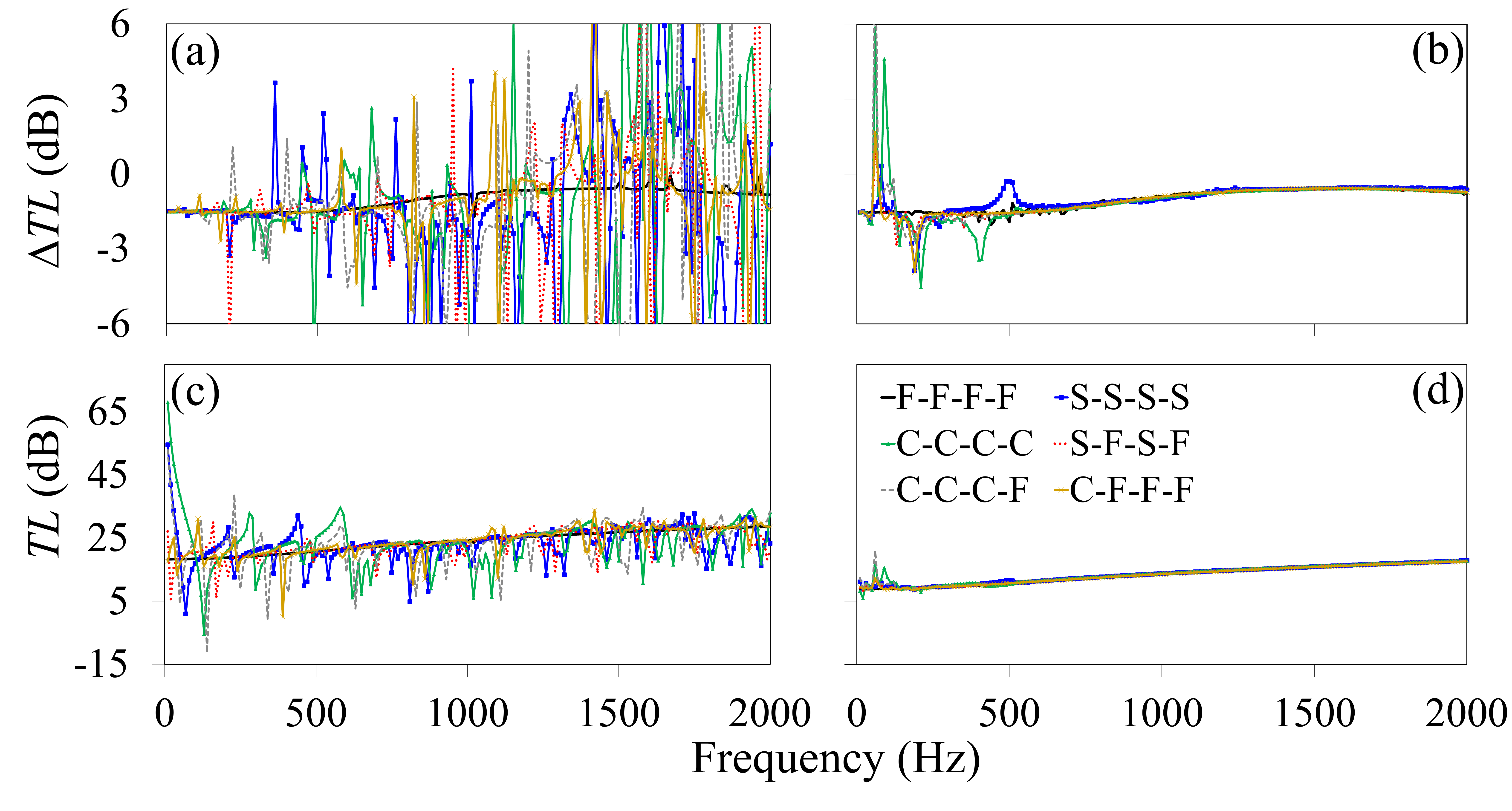}
\caption{(a, b) $\Delta TL$ of a baffled plate with six different boundary conditions between the excitation of DAF and obliquely incident plane wave with $(\theta_{\rm in}, \varphi_{\rm in})=(45^\circ, 0^\circ)$; (c, d) $TL$ of the baffled plate under DAF excitation. (a, c) Aluminum plate, (b, d) rubber plate.}
\label{fig:TL_DAF_delta}
\end{figure}

A comparison of TL between the excitation of the obliquely incident plane wave with $(\theta_{\rm in}, \varphi_{\rm in})=(45^\circ, 0^\circ)$ and the DAF is made in Fig. \ref{fig:TL_DAF_delta}. A deviation is defined by $\Delta TL = TL_{\rm DAF} - TL_{\rm plane}$. Clearly, the sound TL under DAF is slightly smaller than that under the obliquely incident plane wave. As the frequency increases, the overall discrepancy approaches zero. Although there exist many discrepancies at some of the valley locations in the high frequency range of the spectrum for the aluminum plate whose damping is small ($\eta = 0.001$), the general trends are similar to that of the F-F-F-F plate (Fig. \ref{fig:TL_DAF_delta}a). When the material is changed to engineering rubber ($\rho = 370\ {\rm kg\ m^{-3}}$, $E_0 = 2.3\ \rm MPa$, $\mu = 0.4$, $\eta =0.1$, $h= 2\ \rm mm$), the high-frequency peaks and valleys disappear (Fig. \ref{fig:TL_DAF_delta}b). The $\Delta TL$ of the six boundary conditions almost overlap. Moreover, comparisons in Fig. \ref{fig:TL_DAF_delta}(c, d) show that boundary conditions mainly affect the TL of the aluminum plate at low frequencies but the impact is negligible throughout the frequency range for a rubber plate. In this case, a simply supported boundary condition could be used for design change and optimizations of rubber plate or rubber layer.

%%---------------------------%%%%%%%%%%%%%%%%Section 5%%%%%%%%%%%%%%%
\section{Conclusions}
\label{sec:conclusions}

A semi-analytical method using sine series as auxiliary function is proposed for flexural vibration and sound radiation of a thin rectangular plate with general elastic edges. The inspiration of the formulation for vibrating displacement is the use of Fourier sine series as auxiliary coefficients of cosine functions that are trial functions. Rayleigh-Ritz method is employed to find solutions of the vibration. The proposed method can approximately predict the transversely vibrating frequencies and mode shapes of a thin rectangular plate with elastic edges, validated by literature results of 21 cases involving classical boundary conditions (i.e., combinations of free, simply supported, and clamped ends). It is able to predict the vibrating response and sound radiation under baffled or unbaffled conditions, and under any excitation of point force, obliquely incident plane wave, or DAF, which is verified by experimental results, FE-BEM and FE-SEA simulations.

A post-process program is developed for effectively computing the vibrating response and quadruple integrals of sound radiation impedances involving high-order modes. The formulation of vibrating displacement incorporates three items of two-dimensional sine and/or cosine series. As a result, firstly, five tensor products of second-order partial derivatives of the mode function vector $\mathbf{\Gamma}$ are derived by providing four constant vectors (i.e., $\mathbf{q}_M^{}$, $\mathbf{q}_N^{}$, $\mathbf{1}_M^{}$, and $\mathbf{1}_N^{}$), which is conducive to obtaining mass, stiffness and radiation impedance matrices. Secondly, the quadruple integral calculating radiation impedance matrices can be reduced to a twofold integral, whose integrands have closed solutions. Derived formulas for modal forces of point excitation, incident plane wave, and DAF are also presented in this paper. Based on these formulas, a Python package is developed for computing natural frequencies, mode shapes, vibrating response, sound radiation, and TL. High-order modes, up to 5 kHz, are efficiently and accurately predicted. So is the TL under obliquely incident plane wave or DAF.

A comparison is made for different elastic boundary conditions of thin rectangular plates. The effect of boundary conditions on the sound TL at high frequencies (e.g., $\textgreater$ 1 kHz for current study) is found to be negligibly small, whether under obliquely incident acoustic excitation or DAF excitation. The TL under DAF excitation is slightly smaller than under obliquely incident plane wave, but they are very close and the difference approaches zero with the increase of frequency.

%%---------------------------%%%%%%%%%%%%%%%% Acknowledgements %%%%%%%%%%%%%%%
\section*{Acknowledgements}
\label{sec:Acknowledgements}
\addcontentsline{toc}{section}{Acknowledgements}
This work is funded by the National Natural Science Foundation of China (12302359). Authors are grateful for valuable suggestions given by Prof Yu Zhou from Center for Turbulence Control, Harbin Institute of Technology (Shenzhen).
\appendix

%%---------------------------%%%%%%%%%%%%%%%% Appendix A %%%%%%%%%%%%%%%
%%---------------------------------------------------------
\section{The expressions used in Eq. (\ref{eq:DisVector})}
\label{sec:A1}
\setcounter{equation}{0}

The mode function vector $\mathbf{\Gamma}(x,y)$ and the amplitude vector $\mathbf{a}$ are as follows:

\renewcommand{\theequation}{A.\arabic{equation}}
\begin{equation}\label{eq:A-1}
\begin{aligned}
\mathbf{\Gamma}(x,y)&=\Bigl\{{\rm cos}(\lambda_0^{L_x}x){\rm cos}(\lambda_0^{L_y}y), {\rm cos}(\lambda_0^{L_x}x){\rm cos}(\lambda_1^{L_y}y), {\rm cos}(\lambda_0^{L_x}x){\rm cos}(\lambda_2^{L_y}y), \cdots, \\
&{\rm cos}(\lambda_0^{L_x}x){\rm cos}(\lambda_{N}^{L_y}y), {\rm cos}(\lambda_1^{L_x}x){\rm cos}(\lambda_0^{L_y}y), \cdots, {\rm cos}(\lambda_1^{L_x}x){\rm cos}(\lambda_{N}^{L_y}y), \cdots, \\
&\cdots, {\rm cos}(\lambda_{M}^{L_x}x){\rm cos}(\lambda_{N}^{L_y}y), {\rm cos}(\lambda_{0}^{L_x}x){\rm sin}(\lambda_{0}^{L_y}y), \cdots, \cdots, \\
&{\rm cos}(\lambda_{M}^{L_x}x){\rm sin}(\lambda_{N}^{L_y}y), {\rm sin}(\lambda_{0}^{L_x}x){\rm cos}(\lambda_{0}^{L_y}y), \cdots, \cdots, {\rm sin}(\lambda_{M}^{L_x}x){\rm cos}(\lambda_{N}^{L_y}y)\Bigr\}^T,
\end{aligned}
\end{equation}

\begin{equation}\label{eq:A-2}
\begin{aligned}
\mathbf{a}=\Bigl\{&a_{00}, a_{01}, a_{02}, \cdots, a_{0N}, a_{10}, \cdots, a_{1N}, \cdots, \cdots, a_{MN}, b_{00}, b_{01}, b_{02}, \cdots, \cdots, \\
&b_{MN}, c_{00}, c_{01}, c_{02}, \cdots, \cdots, c_{MN} \Bigr\}^T.
\end{aligned}
\end{equation}

%%---------------------------%%%%%%%%%%%%%%%% Appendix B %%%%%%%%%%%%%%%
%%---------------------------------------------------------
\section{Matrices of $\mathbf{A}$ and $\mathbf{B}$ used in Eqs. (\ref{eq:Gammaxyxy})} 
\label{sec:B1}
\setcounter{equation}{0}

The coefficient matrix $\mathbf{A}$ has the following form,

\renewcommand{\theequation}{B.\arabic{equation}}
\begin{equation}\label{eq:B-1}
\mathbf{A}=\begin{bmatrix} \mathbf{A}_{0} & \mathbf{A}_{0} & \mathbf{A}_{0} \\
\mathbf{A}_{0}^T & \mathbf{A}_{0} & \mathbf{A}_{0} \\
\mathbf{A}_{0}^T & \mathbf{A}_{0}^T & \mathbf{A}_{0} \\
\end{bmatrix},
\end{equation}
with

\begin{equation}\label{eq:B-2}
\mathbf{A}_0 = \bigl(\mathbf{q}_M^{} \otimes \mathbf{q}_N^{}\bigr) \bigl(\mathbf{1}_M^{} \otimes \mathbf{1}_N^{}\bigr)^T.
\end{equation}

To write the matrix $\mathbf{B}$, we use the two symbolic formulas as follows:

\begin{equation}\label{eq:B-3}
\bigl[\mathbf{1}\bigr]_{(M+1),(M+1)} = \mathbf{1}_M^{}\mathbf{1}_M^{T},
\end{equation}

\begin{equation}\label{eq:B-4}
\bigl[\mathbf{1}\bigr]_{(M+1)(N+1),(M+1)(N+1)} = \bigl(\mathbf{1}_M^{}\otimes\mathbf{1}_N^{}\bigr)\bigl(\mathbf{1}_M^{}\otimes\mathbf{1}_N^{}\bigr)^{T} = \bigl(\mathbf{1}_M^{}\mathbf{1}_M^{T}\bigr)\otimes \bigl(\mathbf{1}_N^{}\mathbf{1}_N^{T}\bigr).
\end{equation}
Then, the coefficient matrix $\mathbf{B}$ can be expressed as

\begin{equation}\label{eq:B-5}
\mathbf{B} = \begin{bmatrix} \bigl[\mathbf{1}\bigr]_{(M+1)(N+1),(M+1)(N+1)} & \bigl[\mathbf{1}\bigr]_{(M+1)(N+1),(M+1)(N+1)} & \bigl[\mathbf{1}\bigr]_{(M+1)(N+1),(M+1)(N+1)} \\
\bigl[\mathbf{1}\bigr]_{(M+1)(N+1),(M+1)(N+1)} & \bigl[\mathbf{1}\bigr]_{(M+1)(N+1),(M+1)(N+1)} & \bigl[\mathbf{1}\bigr]_{(M+1),(M+1)} \otimes \mathbf{C} \\
\bigl[\mathbf{1}\bigr]_{(M+1)(N+1),(M+1)(N+1)} & \bigl[\mathbf{1}\bigr]_{(M+1),(M+1)} \otimes \mathbf{C}^T & \bigl[\mathbf{1}\bigr]_{(M+1)(N+1),(M+1)(N+1)} \\ \end{bmatrix},
\end{equation}
with

\begin{equation}\label{eq:B-6}
\mathbf{C} = \begin{bmatrix} 
0 & 0 & 0 & \cdots & 0 \\
0 & (1/1)^2 & (2/1)^2 & \cdots & (N/1)^2 \\
0 & (1/2)^2 & (2/2)^2 & \cdots & (N/2)^2 \\ 
\vdots & \vdots & \vdots & \ddots & \vdots \\
0 & (1/N)^2 & (2/N)^2 & \cdots & (N/N)^2 \\ 
\end{bmatrix}.
\end{equation}

%%---------------------------%%%%%%%%%%%%%%%% Appendix C %%%%%%%%%%%%%%%
%%---------------------------------------------------------
\section{Integrals for calculation of mass and stiffness matrices} 
\label{sec:C1}
\setcounter{equation}{0}

The following three integrals are used for the calculation of $\iint_{S_0}\bm\Gamma\bm\Gamma^T{\rm d}S$, $\iint_{S_0}\bm\Gamma_x\bm\Gamma_x^T{\rm d}S$, and $\iint_{S_0}\bm\Gamma_y\bm\Gamma_y^T{\rm d}S$, where $S_0$ indicates the area of the plate.

\renewcommand{\theequation}{C.\arabic{equation}}
\begin{equation}\label{eq:Integral3}
\begin{aligned}
\int_0^{L_x} {\rm sin}(\lambda_m^{L_x}x){\rm sin}(\lambda_s^{L_x}x){\rm d}x = 
\begin{cases} 
L_x/2 & m=s \neq 0 \\ 
0 & {\rm else} \\
\end{cases},
\end{aligned}
\end{equation} 

\begin{equation}\label{eq:Integral4}
\begin{aligned}
\int_0^{L_x} {\rm cos}(\lambda_m^{L_x}x){\rm cos}(\lambda_s^{L_x}x){\rm d}x = 
\begin{cases} 
L_x/2 & m=s \neq 0 \\
L_x & m=s=0 \\ 
0 & {\rm else} \\
\end{cases},
\end{aligned}
\end{equation} 

\begin{equation}\label{eq:Integral5}
\begin{aligned}
\int_0^{L_x} {\rm sin}(\lambda_m^{L_x}x){\rm cos}(\lambda_s^{L_x}x){\rm d}x = 
\begin{cases} 
L_x/2 & m=s \\ 
L_x\frac{m\bigl((-1)^{m+s}-1\bigr)}{\pi\bigl(s^2-m^2\bigr)} & m \neq s \\
\end{cases}.
\end{aligned}
\end{equation} 

%%---------------------------%%%%%%%%%%%%%%%% Appendix D %%%%%%%%%%%%%%%
%%---------------------------------------------------------
\section{Integrals used in Table \ref{tab:correspondance}} 
\label{sec:D1}

The closed solutions of the 4 integrals, $H_{kl}^{\rm cc}(\alpha)$, $H_{kl}^{\rm cs}(\alpha)$, $H_{kl}^{\rm sc}(\alpha)$, and $H_{kl}^{\rm ss}(\alpha)$, are as follows:

\setcounter{equation}{0}
\renewcommand{\theequation}{D.\arabic{equation}}
\begin{equation}\label{eq:D-1}
\begin{aligned}
H_{kl}^{\rm cc}(\alpha)& = \int_0^{1-\alpha}{\rm cos}(k\pi(\alpha+\beta)){\rm cos}(l\pi\beta){\rm d}\beta \\
&= \begin{cases} 
\frac{-k{\rm sin}(k\pi\alpha)+ l(-1)^k{\rm sin}(l\pi(\alpha-1))}{(k^2-l^2)\pi} & k \neq l \\ 
-\frac{2(\alpha-1)l\pi{\rm cos}(l\pi\alpha)+{\rm sin}(l\pi(\alpha-2))+{\rm sin}(l\pi\alpha)}{4l\pi} & k=l \neq 0 \\
1-\alpha & k=l=0 \\
\end{cases},
\end{aligned}
\end{equation}

\begin{equation}\label{eq:D-2}
\begin{aligned}
H_{kl}^{\rm cs}(\alpha)& = \int_0^{1-\alpha}{\rm cos}(k\pi(\alpha+\beta)){\rm sin}(l\pi\beta){\rm d}\beta \\
&= \begin{cases} 
-l\frac{{\rm cos}(k\pi\alpha)+(-1)^{1+k}{\rm cos}(l\pi(\alpha-1))}{(k^2-l^2)\pi} & k \neq l \\ 
\frac{2(\alpha-1)l\pi{\rm sin}(l\pi\alpha)-{\rm cos}(l\pi(\alpha-2))+{\rm cos}(l\pi\alpha)}{4l\pi} & k=l \neq 0 \\
0 & k=l=0 \\
\end{cases},
\end{aligned}
\end{equation}

\begin{equation}\label{eq:D-3}
\begin{aligned}
H_{kl}^{\rm sc}(\alpha)& = \int_0^{1-\alpha}{\rm sin}(k\pi(\alpha+\beta)){\rm cos}(l\pi\beta){\rm d}\beta \\
&= \begin{cases} 
k\frac{{\rm cos}(k\pi\alpha)+(-1)^{1+k}{\rm cos}(l\pi(\alpha-1))}{(k^2-l^2)\pi} & k \neq l \\ 
\frac{-2(\alpha-1)l\pi{\rm sin}(l\pi\alpha)-{\rm cos}(l\pi(\alpha-2))+{\rm cos}(l\pi\alpha)}{4l\pi} & k=l \neq 0 \\
0 & k=l=0 \\
\end{cases},
\end{aligned}
\end{equation}

\begin{equation}\label{eq:D-4}
\begin{aligned}
H_{kl}^{\rm ss}(\alpha)& = \int_0^{1-\alpha}{\rm sin}(k\pi(\alpha+\beta)){\rm sin}(l\pi\beta){\rm d}\beta \\
&= \begin{cases} 
\frac{-l{\rm sin}(k\pi\alpha)+ k(-1)^k{\rm sin}(l\pi(\alpha-1))}{(k^2-l^2)\pi} & k \neq l \\ 
\frac{-2(\alpha-1)l\pi{\rm cos}(l\pi\alpha)+{\rm sin}(l\pi(\alpha-2))+{\rm sin}(l\pi\alpha)}{4l\pi} & k=l \neq 0 \\
0 & k=l=0 \\
\end{cases}.
\end{aligned}
\end{equation}

%%---------------------------%%%%%%%%%%%%%%%% Appendix E %%%%%%%%%%%%%%%
%%---------------------------------------------------------
\section{Expressions used in Eqs. (\ref{eq:fbl1-3})} 
\label{sec:E1}
\setcounter{equation}{0}

The closed solution of the integral $Y_{L_x,m}^{\rm c}$ is as

\renewcommand{\theequation}{E.\arabic{equation}}
\begin{equation}\label{eq:E-1}
\begin{aligned}
Y_{L_x,m}^{\rm c}& = \int_0^{L_x}{\rm exp}(-{\rm j}k_0 x{\rm sin}\theta_{\rm in}{\rm cos}\varphi_{\rm in}){\rm cos}(\lambda_m^{L_x}x){\rm d}x \\
&= \begin{cases} 
\frac{L_x}{2}{\rm sgn}({\rm sin}\theta_{\rm in}{\rm cos}\varphi_{\rm in}) & (m\pi)^2=(k_0 L_x{\rm sin}\theta_{\rm in}{\rm cos}\varphi_{\rm in})^2 \neq 0 \\ 
\frac{{\rm j}k_0 L_x^2 {\rm sin}\theta_{\rm in}{\rm cos}\varphi_{\rm in}\bigl[1-(-1)^m {\rm exp}(-{\rm j}k_0 L_x {\rm sin}\theta_{\rm in}{\rm cos}\varphi_{\rm in})\bigr]}{(m\pi)^2 - (k_0 L_x{\rm sin}\theta_{\rm in}{\rm cos}\varphi_{\rm in})^2} & (m\pi)^2 \neq (k_0 L_x{\rm sin}\theta_{\rm in}{\rm cos}\varphi_{\rm in})^2 \\
L_x & m={\rm sin}\theta_{\rm in}{\rm cos}\varphi_{\rm in}=0 \\
\end{cases},
\end{aligned}
\end{equation}
where ${\rm sgn}(x)=1$ for $x>0$; ${\rm sgn}(x)=-1$ for $x<0$; ${\rm sgn}(0)=0$. Towards the integral $Y_{L_y,n}^{\rm c}$, replace $L_x$, $m$ and ${\rm cos}\varphi_{\rm in}$ in Eq. (\ref{eq:E-1}) with $L_y$, $n$ and ${\rm sin}\varphi_{\rm in}$, respectively.

The closed solution of the integral $Y_{L_x,m}^{\rm s}$ is as

\begin{equation}\label{eq:E-2}
\begin{aligned}
Y_{L_x,m}^{\rm s}& = \int_0^{L_x}{\rm exp}(-{\rm j}k_0 x{\rm sin}\theta_{\rm in}{\rm cos}\varphi_{\rm in}){\rm sin}(\lambda_m^{L_x}x){\rm d}x \\
&= \begin{cases} 
-\frac{{\rm j}L_x}{2}{\rm sgn}({\rm sin}\theta_{\rm in}{\rm cos}\varphi_{\rm in}) & (m\pi)^2=(k_0 L_x{\rm sin}\theta_{\rm in}{\rm cos}\varphi_{\rm in})^2 \\ 
\frac{m\pi L_x \bigl[1-(-1)^m {\rm exp}(-{\rm j}k_0 L_x {\rm sin}\theta_{\rm in}{\rm cos}\varphi_{\rm in})\bigr]}{(m\pi)^2 - (k_0 L_x{\rm sin}\theta_{\rm in}{\rm cos}\varphi_{\rm in})^2} & (m\pi)^2 \neq (k_0 L_x{\rm sin}\theta_{\rm in}{\rm cos}\varphi_{\rm in})^2 \\
\end{cases}.
\end{aligned}
\end{equation}
Towards the integral $Y_{L_y,n}^{\rm s}$, replace $L_x$, $m$ and ${\rm cos}\varphi_{\rm in}$ in Eq. (\ref{eq:E-2}) with $L_y$, $n$ and ${\rm sin}\varphi_{\rm in}$, respectively.

%\bibliographystyle{unsrt}
\bibliographystyle{elsarticle-num-names} 
\bibliography{Refs_file}

\begin{thebibliography}{63}
\expandafter\ifx\csname natexlab\endcsname\relax\def\natexlab#1{#1}\fi
\providecommand{\url}[1]{\texttt{#1}}
\providecommand{\href}[2]{#2}
\providecommand{\path}[1]{#1}
\providecommand{\DOIprefix}{doi:}
\providecommand{\ArXivprefix}{arXiv:}
\providecommand{\URLprefix}{URL: }
\providecommand{\Pubmedprefix}{pmid:}
\providecommand{\doi}[1]{\href{http://dx.doi.org/#1}{\path{#1}}}
\providecommand{\Pubmed}[1]{\href{pmid:#1}{\path{#1}}}
\providecommand{\bibinfo}[2]{#2}
\ifx\xfnm\relax \def\xfnm[#1]{\unskip,\space#1}\fi
%Type = Article
\bibitem[{Kim and Park(2020)}]{Kim2020}
\bibinfo{author}{D.~Kim}, \bibinfo{author}{N.~C. Park},
\newblock \bibinfo{title}{Calculation and reduction of sound radiation from a
  thin plate structure excited by complex inputs},
\newblock \bibinfo{journal}{J. Sound Vib.} \bibinfo{volume}{484}
  (\bibinfo{year}{2020}) \bibinfo{pages}{115517}.
  \DOIprefix\doi{10.1016/j.jsv.2020.115517}.
%Type = Article
\bibitem[{Shao et~al.(2021)Shao, Luo, Deng, Zeng, Yang, Wu, and Jin}]{Shao2021}
\bibinfo{author}{J.~Shao}, \bibinfo{author}{Q.~Luo}, \bibinfo{author}{G.~Deng},
  \bibinfo{author}{T.~Zeng}, \bibinfo{author}{J.~Yang},
  \bibinfo{author}{X.~Wu}, \bibinfo{author}{C.~Jin},
\newblock \bibinfo{title}{Experimental study on influence of wall acoustic
  materials of 3d cavity for targeted energy transfer of a nonlinear membrane
  absorber},
\newblock \bibinfo{journal}{Appl. Acoust.} \bibinfo{volume}{184}
  (\bibinfo{year}{2021}) \bibinfo{pages}{108342}.
  \DOIprefix\doi{10.1016/j.apacoust.2021.108342}.
%Type = Article
\bibitem[{Shao et~al.(2023)Shao, Zhao, Yang, Luo, and Wu}]{Shao2023}
\bibinfo{author}{J.~Shao}, \bibinfo{author}{H.~Zhao},
  \bibinfo{author}{J.~Yang}, \bibinfo{author}{Q.~Luo}, \bibinfo{author}{X.~Wu},
\newblock \bibinfo{title}{Research on suppressing radiation noise of plate
  inside acoustic cavity based on targeted energy transfer of nonlinear energy
  sink},
\newblock \bibinfo{journal}{J. Braz. Soc. Mech. Sci.} \bibinfo{volume}{45}
  (\bibinfo{year}{2023}) \bibinfo{pages}{220}.
  \DOIprefix\doi{10.1007/s40430-023-04109-w}.
%Type = Article
\bibitem[{Marchetto et~al.(2017)Marchetto, Maxit, Robin, and
  Berry}]{Marchetto2017}
\bibinfo{author}{C.~Marchetto}, \bibinfo{author}{L.~Maxit},
  \bibinfo{author}{O.~Robin}, \bibinfo{author}{A.~Berry},
\newblock \bibinfo{title}{Vibroacoustic response of panels under diffuse
  acoustic field excitation from sensitivity functions and reciprocity
  principles},
\newblock \bibinfo{journal}{J. Acoust. Soc. Am.} \bibinfo{volume}{141}
  (\bibinfo{year}{2017}) \bibinfo{pages}{4508--4521}.
  \DOIprefix\doi{10.1121/1.4985126}.
%Type = Article
\bibitem[{Deng et~al.(2021)Deng, Shao, Zheng, and Wu}]{Deng2021a}
\bibinfo{author}{G.~Deng}, \bibinfo{author}{J.~Shao},
  \bibinfo{author}{S.~Zheng}, \bibinfo{author}{X.~Wu},
\newblock \bibinfo{title}{Study on sound transmission loss modeling through
  simplified sealing specimens and an automotive door sealing system},
\newblock \bibinfo{journal}{Noise Control Eng. J.} \bibinfo{volume}{69}
  (\bibinfo{year}{2021}) \bibinfo{pages}{301--330}.
  \DOIprefix\doi{10.3397/1/376929}.
%Type = Article
\bibitem[{Bonness et~al.(2017)Bonness, Fahnline, Lysak, and
  Shepherd}]{Bonness2017}
\bibinfo{author}{W.~K. Bonness}, \bibinfo{author}{J.~B. Fahnline},
  \bibinfo{author}{P.~D. Lysak}, \bibinfo{author}{M.~R. Shepherd},
\newblock \bibinfo{title}{Modal forcing functions for structural vibration from
  turbulent boundary layer flow},
\newblock \bibinfo{journal}{J. Sound Vib.} \bibinfo{volume}{395}
  (\bibinfo{year}{2017}) \bibinfo{pages}{224--239}.
  \DOIprefix\doi{10.1016/j.jsv.2017.02.023}.
%Type = Article
\bibitem[{Hambric et~al.(2004)Hambric, Hwang, and Bonness}]{Hambric2004}
\bibinfo{author}{S.~A. Hambric}, \bibinfo{author}{Y.~F. Hwang},
  \bibinfo{author}{W.~K. Bonness},
\newblock \bibinfo{title}{Vibrations of plates with clamped and free edges
  excited by low-speed turbulent boundary layer flow},
\newblock \bibinfo{journal}{J. Fluids Struct.} \bibinfo{volume}{19}
  (\bibinfo{year}{2004}) \bibinfo{pages}{93--110}.
  \DOIprefix\doi{10.1016/j.jfluidstructs.2003.09.002}.
%Type = Article
\bibitem[{Shao et~al.(2020)Shao, Yang, Wu, Wang, and Deng}]{Shao2020}
\bibinfo{author}{J.~Shao}, \bibinfo{author}{J.~Yang}, \bibinfo{author}{X.~Wu},
  \bibinfo{author}{C.~Wang}, \bibinfo{author}{G.~Deng},
\newblock \bibinfo{title}{Study on radiated noise of a panel under fluctuating
  surface pressure due to an idealized side mirror},
\newblock \bibinfo{journal}{Appl. Sci.} \bibinfo{volume}{10}
  (\bibinfo{year}{2020}) \bibinfo{pages}{994}.
  \DOIprefix\doi{10.3390/app10030994}.
%Type = Article
\bibitem[{Hasan(2023)}]{Hasan2023}
\bibinfo{author}{M.~Z. Hasan},
\newblock \bibinfo{title}{An experimental study on the sound transmission loss
  of dissimilar fuselage sandwich panels of turbojet aircraft},
\newblock \bibinfo{journal}{Thin-Walled Struct.} \bibinfo{volume}{184}
  (\bibinfo{year}{2023}) \bibinfo{pages}{110417}.
  \DOIprefix\doi{10.1016/j.tws.2022.110417}.
%Type = Article
\bibitem[{Ghiringhelli et~al.(2013)Ghiringhelli, Terraneo, and
  Vigoni}]{Ghiringhelli2013}
\bibinfo{author}{G.~Ghiringhelli}, \bibinfo{author}{M.~Terraneo},
  \bibinfo{author}{E.~Vigoni},
\newblock \bibinfo{title}{Improvement of structures vibroacoustics by
  widespread embodiment of viscoelastic materials},
\newblock \bibinfo{journal}{Aerosp. Sci. Technol.} \bibinfo{volume}{28}
  (\bibinfo{year}{2013}) \bibinfo{pages}{227--241}.
  \DOIprefix\doi{10.1016/j.ast.2012.11.003}.
%Type = Article
\bibitem[{He et~al.(2021)He, Wan, Liu, Wen, and Yang}]{He2021}
\bibinfo{author}{Y.~He}, \bibinfo{author}{R.~Wan}, \bibinfo{author}{Y.~Liu},
  \bibinfo{author}{S.~Wen}, \bibinfo{author}{Z.~Yang},
\newblock \bibinfo{title}{Transmission characteristics and mechanism study of
  hydrodynamic and acoustic pressure through a side window of drivaer model
  based on modal analytical approach},
\newblock \bibinfo{journal}{J. Sound Vib.} \bibinfo{volume}{501}
  (\bibinfo{year}{2021}) \bibinfo{pages}{116058}.
  \DOIprefix\doi{10.1016/j.jsv.2021.116058}.
%Type = Article
\bibitem[{Thekinen and Datta(2019)}]{Thekinen201973}
\bibinfo{author}{J.~Thekinen}, \bibinfo{author}{N.~Datta},
\newblock \bibinfo{title}{Rayleigh-ritz method-based analysis of dry coupled
  horizontal-torsional-warping vibration of rectelliptic open-section
  containership bare-hulls},
\newblock \bibinfo{journal}{Appl. Ocean Res.} \bibinfo{volume}{86}
  (\bibinfo{year}{2019}) \bibinfo{pages}{73--86}.
  \DOIprefix\doi{10.1016/j.apor.2019.01.032}.
%Type = Article
\bibitem[{Sehgal and Kumar(2016)}]{Sehgal2016}
\bibinfo{author}{S.~Sehgal}, \bibinfo{author}{H.~Kumar},
\newblock \bibinfo{title}{Structural dynamic model updating techniques: A state
  of the art review},
\newblock \bibinfo{journal}{Arch. Computat. Methods Eng.} \bibinfo{volume}{23}
  (\bibinfo{year}{2016}) \bibinfo{pages}{515--533}.
  \DOIprefix\doi{10.1007/s11831-015-9150-3}.
%Type = Article
\bibitem[{Chen and Hwu(2014)}]{CHEN201422}
\bibinfo{author}{Y.~Chen}, \bibinfo{author}{C.~Hwu},
\newblock \bibinfo{title}{Boundary element method for vibration analysis of
  two-dimensional anisotropic elastic solids containing holes, cracks or
  interfaces},
\newblock \bibinfo{journal}{Eng. Anal. Bound. Elem.} \bibinfo{volume}{40}
  (\bibinfo{year}{2014}) \bibinfo{pages}{22--35}.
  \DOIprefix\doi{10.1016/j.enganabound.2013.11.013}.
%Type = Article
\bibitem[{Zhang et~al.(2023)Zhang, Li, Hao, Lin, Li, and Su}]{ZHANG2023110630}
\bibinfo{author}{Y.~Zhang}, \bibinfo{author}{Z.~Li}, \bibinfo{author}{R.~Hao},
  \bibinfo{author}{W.~Lin}, \bibinfo{author}{L.~Li}, \bibinfo{author}{D.~Su},
\newblock \bibinfo{title}{High-fidelity time-series data synthesis based on
  finite element simulation and data space mapping},
\newblock \bibinfo{journal}{Mech. Syst. Signal. Process.} \bibinfo{volume}{200}
  (\bibinfo{year}{2023}) \bibinfo{pages}{110630}.
  \DOIprefix\doi{10.1016/j.ymssp.2023.110630}.
%Type = Article
\bibitem[{Deng et~al.(2020)Deng, Shao, Zheng, and Wu}]{Deng2020}
\bibinfo{author}{G.~Deng}, \bibinfo{author}{J.~Shao},
  \bibinfo{author}{S.~Zheng}, \bibinfo{author}{X.~Wu},
\newblock \bibinfo{title}{Optimal study on sectional geometry of rubber layers
  and cavities based on the vibro-acoustic coupling model with a sine-auxiliary
  function},
\newblock \bibinfo{journal}{Appl. Acoust.} \bibinfo{volume}{170}
  (\bibinfo{year}{2020}) \bibinfo{pages}{107522}.
  \DOIprefix\doi{10.1016/j.apacoust.2020.107522}.
%Type = Article
\bibitem[{Deng et~al.(2021)Deng, Zheng, Shao, Wu, and Chen}]{Deng2021b}
\bibinfo{author}{G.~Deng}, \bibinfo{author}{S.~Zheng},
  \bibinfo{author}{J.~Shao}, \bibinfo{author}{X.~Wu},
  \bibinfo{author}{Z.~Chen},
\newblock \bibinfo{title}{Structural-acoustic coupling analytical model and
  sound insulation optimization of rubber layers and cavity},
\newblock \bibinfo{journal}{J. Tongji Univ. Nat. Sci} \bibinfo{volume}{49}
  (\bibinfo{year}{2021}) \bibinfo{pages}{280 -- 288}.
  \DOIprefix\doi{10.11908/j.issn.0253-374x.20381}.
%Type = Article
\bibitem[{Liu et~al.(2024)Liu, Cui, Wang, and Wang}]{Liu2024}
\bibinfo{author}{C.~Liu}, \bibinfo{author}{Y.~J. Cui}, \bibinfo{author}{K.~F.
  Wang}, \bibinfo{author}{B.~L. Wang},
\newblock \bibinfo{title}{Fatigue life prediction and energy conversion
  efficiency evaluation of a photovoltaic-thermoelectric device subjected to
  time-varying thermal and wind hybrid loads},
\newblock \bibinfo{journal}{Int. J. Solids. Struct.} \bibinfo{volume}{293}
  (\bibinfo{year}{2024}) \bibinfo{pages}{112741}.
  \DOIprefix\doi{10.1016/j.ijsolstr.2024.112741}.
%Type = Article
\bibitem[{Liu et~al.(2023{\natexlab{a}})Liu, Cui, Wang, and Wang}]{Liu2023a}
\bibinfo{author}{C.~Liu}, \bibinfo{author}{Y.~J. Cui}, \bibinfo{author}{K.~F.
  Wang}, \bibinfo{author}{B.~L. Wang},
\newblock \bibinfo{title}{Interlaminar mechanical performance of a
  multi-layered device},
\newblock \bibinfo{journal}{Appl. Math. Model} \bibinfo{volume}{122}
  (\bibinfo{year}{2023}{\natexlab{a}}) \bibinfo{pages}{242--264}.
  \DOIprefix\doi{10.1016/j.apm.2023.05.040}.
%Type = Article
\bibitem[{Liu et~al.(2023{\natexlab{b}})Liu, Cui, Wang, and Wang}]{Liu2023b}
\bibinfo{author}{C.~Liu}, \bibinfo{author}{Y.~J. Cui}, \bibinfo{author}{K.~F.
  Wang}, \bibinfo{author}{B.~L. Wang},
\newblock \bibinfo{title}{Bending strength evaluation and power generation
  performance optimization of a curved photovoltaic-thermoelectric hybrid
  device},
\newblock \bibinfo{journal}{Compos. Struct.} \bibinfo{volume}{321}
  (\bibinfo{year}{2023}{\natexlab{b}}) \bibinfo{pages}{117297}.
  \DOIprefix\doi{10.1016/j.compstruct.2023.117297}.
%Type = Article
\bibitem[{Leissa(1973)}]{Leissa1973}
\bibinfo{author}{A.~W. Leissa},
\newblock \bibinfo{title}{The free vibration of rectangular plates},
\newblock \bibinfo{journal}{J. Sound Vib.} \bibinfo{volume}{31}
  (\bibinfo{year}{1973}) \bibinfo{pages}{257--293}.
  \DOIprefix\doi{10.1016/S0022-460X(73)80371-2}.
%Type = Book
\bibitem[{Navier(1819)}]{Navier1819}
\bibinfo{author}{L.~M.~H. Navier}, \bibinfo{title}{Resume des Lecons de
  Mechanique}, \bibinfo{publisher}{Ecole Polytechnique, Paris},
  \bibinfo{year}{1819}.
%Type = Book
\bibitem[{Rayleigh(1945)}]{Rayleigh1945}
\bibinfo{author}{L.~Rayleigh}, \bibinfo{title}{Theory of Sound},
  \bibinfo{publisher}{Dover Publications}, \bibinfo{year}{1945}.
%Type = Book
\bibitem[{Timoshenko(1934)}]{Timoshenko1934}
\bibinfo{author}{S.~Timoshenko}, \bibinfo{title}{Theory of Elasticity},
  \bibinfo{publisher}{McGraw-Hill Book Company, Inc.}, \bibinfo{year}{1934}.
%Type = Article
\bibitem[{Levy(1899)}]{Levy1899}
\bibinfo{author}{M.~Levy},
\newblock \bibinfo{title}{Sur l'equlibre elastique d'une plaque rectangulaire},
\newblock \bibinfo{journal}{C. R. Acad. Sci.} \bibinfo{volume}{129}
  (\bibinfo{year}{1899}) \bibinfo{pages}{535 -- 539}.
%Type = Book
\bibitem[{Voigt(1893)}]{Voigt1893}
\bibinfo{author}{W.~Voigt}, \bibinfo{title}{Bemerkungen zu dem Problem der
  transversalen Schwingungen rechteckiger Platten}, \bibinfo{publisher}{Nachr.
  Ges. Wiss. (Gottingen)}, \bibinfo{year}{1893}.
%Type = Article
\bibitem[{Ritz(1909)}]{Ritz1909}
\bibinfo{author}{W.~Ritz},
\newblock \bibinfo{title}{Theorie der transversalschwingungen einer
  quadratischen platte mit freien randern},
\newblock \bibinfo{journal}{Annalen der Physik} \bibinfo{volume}{333}
  (\bibinfo{year}{1909}) \bibinfo{pages}{737--786}.
  \DOIprefix\doi{10.1002/andp.19093330403}.
%Type = Article
\bibitem[{Pickett(1939)}]{Pickett1939}
\bibinfo{author}{G.~Pickett},
\newblock \bibinfo{title}{{Solution of Rectangular Clamped Plate With Lateral
  Load by Generalized Energy Method}},
\newblock \bibinfo{journal}{J. Appl. Mech} \bibinfo{volume}{6}
  (\bibinfo{year}{1939}) \bibinfo{pages}{A168--A170}.
  \DOIprefix\doi{10.1115/1.4008967}.
%Type = Article
\bibitem[{Young(1950)}]{Young1950}
\bibinfo{author}{D.~Young},
\newblock \bibinfo{title}{{Vibration of Rectangular Plates by the Ritz
  Method}},
\newblock \bibinfo{journal}{J. Appl. Mech} \bibinfo{volume}{17}
  (\bibinfo{year}{1950}) \bibinfo{pages}{448--453}.
  \DOIprefix\doi{10.1115/1.4010175}.
%Type = Article
\bibitem[{Warburton(1954)}]{Warburton1954}
\bibinfo{author}{G.~B. Warburton},
\newblock \bibinfo{title}{The vibration of rectangular plates},
\newblock \bibinfo{journal}{Proceedings of the Institution of Mechanical
  Engineers} \bibinfo{volume}{168} (\bibinfo{year}{1954})
  \bibinfo{pages}{371--384}.
  \DOIprefix\doi{10.1243/PIME\_PROC\_1954\_168\_040\_02}.
%Type = Article
\bibitem[{Dickinson(1978)}]{DICKINSON19781}
\bibinfo{author}{S.~Dickinson},
\newblock \bibinfo{title}{The buckling and frequency of flexural vibration of
  rectangular isotropic and orthotropic plates using rayleigh's method},
\newblock \bibinfo{journal}{J. Sound Vib.} \bibinfo{volume}{61}
  (\bibinfo{year}{1978}) \bibinfo{pages}{1--8}.
  \DOIprefix\doi{10.1016/0022-460X(78)90036-6}.
%Type = Article
\bibitem[{Laura and Grossi(1981)}]{LAURA1981101}
\bibinfo{author}{P.~Laura}, \bibinfo{author}{R.~Grossi},
\newblock \bibinfo{title}{Transverse vibrations of rectangular plates with
  edges elastically restrained against translation and rotation},
\newblock \bibinfo{journal}{J. Sound Vib.} \bibinfo{volume}{75}
  (\bibinfo{year}{1981}) \bibinfo{pages}{101--107}.
  \DOIprefix\doi{10.1016/0022-460X(81)90237-6}.
%Type = Article
\bibitem[{Warburton and Edney(1984)}]{WARBURTON1984537}
\bibinfo{author}{G.~Warburton}, \bibinfo{author}{S.~Edney},
\newblock \bibinfo{title}{Vibrations of rectangular plates with elastically
  restrained edges},
\newblock \bibinfo{journal}{J. Sound Vib.} \bibinfo{volume}{95}
  (\bibinfo{year}{1984}) \bibinfo{pages}{537--552}.
  \DOIprefix\doi{10.1016/0022-460X(84)90236-0}.
%Type = Article
\bibitem[{Bhat(1985)}]{BHAT1985493}
\bibinfo{author}{R.~Bhat},
\newblock \bibinfo{title}{Natural frequencies of rectangular plates using
  characteristic orthogonal polynomials in rayleigh-ritz method},
\newblock \bibinfo{journal}{J. Sound Vib.} \bibinfo{volume}{102}
  (\bibinfo{year}{1985}) \bibinfo{pages}{493--499}.
  \DOIprefix\doi{10.1016/S0022-460X(85)80109-7}.
%Type = Article
\bibitem[{Cupial(1997)}]{CUPIAL1997385}
\bibinfo{author}{P.~Cupial},
\newblock \bibinfo{title}{Calculation of the natural frequencies of composite
  plates by the rayleigh-ritz method with orthogonal polynominals},
\newblock \bibinfo{journal}{J. Sound Vib.} \bibinfo{volume}{201}
  (\bibinfo{year}{1997}) \bibinfo{pages}{385--387}.
  \DOIprefix\doi{10.1006/jsvi.1996.0802}.
%Type = Article
\bibitem[{Dickinson and {Di Blasio}(1986)}]{DICKINSON198651}
\bibinfo{author}{S.~Dickinson}, \bibinfo{author}{A.~{Di Blasio}},
\newblock \bibinfo{title}{On the use of orthogonal polynomials in the
  rayleigh-ritz method for the study of the flexural vibration and buckling of
  isotropic and orthotropic rectangular plates},
\newblock \bibinfo{journal}{J. Sound. Vib.} \bibinfo{volume}{108}
  (\bibinfo{year}{1986}) \bibinfo{pages}{51--62}.
  \DOIprefix\doi{10.1016/S0022-460X(86)80310-8}.
%Type = Article
\bibitem[{Berry et~al.(1990)Berry, Guyader, and Nicolas}]{Berry1990}
\bibinfo{author}{A.~Berry}, \bibinfo{author}{J.~Guyader},
  \bibinfo{author}{J.~Nicolas},
\newblock \bibinfo{title}{A general formulation for the sound radiation from
  rectangular, baffled plates with arbitrary boundary conditions},
\newblock \bibinfo{journal}{J. Acoust. Soc. Am.} \bibinfo{volume}{88}
  (\bibinfo{year}{1990}) \bibinfo{pages}{2792--2802}.
  \DOIprefix\doi{10.1121/1.399682}.
%Type = Article
\bibitem[{Berry(1994)}]{Berry1994}
\bibinfo{author}{A.~Berry},
\newblock \bibinfo{title}{A new formulation for the vibrations and sound
  radiation of fluid-loaded plates with elastic boundary conditions},
\newblock \bibinfo{journal}{J. Acoust. Soc. Am.} \bibinfo{volume}{96}
  (\bibinfo{year}{1994}) \bibinfo{pages}{889--901}.
  \DOIprefix\doi{10.1121/1.410264}.
%Type = Article
\bibitem[{Park et~al.(2003)Park, Siegmund, and Mongeau}]{Park2003}
\bibinfo{author}{J.~Park}, \bibinfo{author}{T.~Siegmund},
  \bibinfo{author}{L.~Mongeau},
\newblock \bibinfo{title}{Analysis of the flow-induced vibrations of
  viscoelastically supported rectangular plates},
\newblock \bibinfo{journal}{J. Sound Vib.} \bibinfo{volume}{261}
  (\bibinfo{year}{2003}) \bibinfo{pages}{225--245}.
  \DOIprefix\doi{10.1016/S0022-460X(02)00955-0}.
%Type = Article
\bibitem[{Park et~al.(2004)Park, Mongeau, and Siegmund}]{Park2004}
\bibinfo{author}{J.~Park}, \bibinfo{author}{L.~Mongeau},
  \bibinfo{author}{T.~Siegmund},
\newblock \bibinfo{title}{An investigation of the flow-induced sound and
  vibration of viscoelastically supported rectangular plates: Experiments and
  model verification},
\newblock \bibinfo{journal}{J. Sound Vib.} \bibinfo{volume}{275}
  (\bibinfo{year}{2004}) \bibinfo{pages}{249--265}.
  \DOIprefix\doi{10.1016/j.jsv.2003.06.017}.
%Type = Article
\bibitem[{Beslin and Nicolas(1997)}]{Beslin1997}
\bibinfo{author}{O.~Beslin}, \bibinfo{author}{J.~Nicolas},
\newblock \bibinfo{title}{A hierarchical functions set for predicting very high
  order plate bending modes with any boundary conditions},
\newblock \bibinfo{journal}{J. Sound Vib.} \bibinfo{volume}{202}
  (\bibinfo{year}{1997}) \bibinfo{pages}{633--655}.
  \DOIprefix\doi{10.1006/jsvi.1996.0797}.
%Type = Article
\bibitem[{Li et~al.(2009)Li, Zhang, Du, and Liu}]{Li2009}
\bibinfo{author}{W.~L. Li}, \bibinfo{author}{X.~Zhang},
  \bibinfo{author}{J.~Du}, \bibinfo{author}{Z.~Liu},
\newblock \bibinfo{title}{An exact series solution for the transverse vibration
  of rectangular plates with general elastic boundary supports},
\newblock \bibinfo{journal}{J. Sound Vib.} \bibinfo{volume}{321}
  (\bibinfo{year}{2009}) \bibinfo{pages}{254--269}.
  \DOIprefix\doi{10.1016/j.jsv.2008.09.035}.
%Type = Article
\bibitem[{Mohammadesmaeili et~al.(2021)Mohammadesmaeili, Motaghian, and
  Mofid}]{MOHAMMADESMAEILI2021104274}
\bibinfo{author}{R.~Mohammadesmaeili}, \bibinfo{author}{S.~Motaghian},
  \bibinfo{author}{M.~Mofid},
\newblock \bibinfo{title}{An innovative series solution for dynamic response of
  rectangular mindlin plate on two-parameter elastic foundation, with general
  boundary conditions},
\newblock \bibinfo{journal}{Eur. J. Mech. A-Solid.} \bibinfo{volume}{88}
  (\bibinfo{year}{2021}) \bibinfo{pages}{104274}.
  \DOIprefix\doi{https://doi.org/10.1016/j.euromechsol.2021.104274}.
%Type = Article
\bibitem[{Wu et~al.(2022)Wu, Chen, and Qu}]{Wu2022}
\bibinfo{author}{T.~Wu}, \bibinfo{author}{Z.~B. Chen}, \bibinfo{author}{J.~J.
  Qu},
\newblock \bibinfo{title}{A modified fourier-ritz method for free vibration of
  rectangular plates with elastic constrains},
\newblock \bibinfo{journal}{Journal of Theoretical and Applied Mechanics}
  \bibinfo{volume}{60} (\bibinfo{year}{2022}) \bibinfo{pages}{77--89}.
  \DOIprefix\doi{10.15632/jtam-pl/144462}.
%Type = Article
\bibitem[{Zhang et~al.(2017)Zhang, Shi, and Wang}]{ZHANG20171}
\bibinfo{author}{H.~Zhang}, \bibinfo{author}{D.~Shi},
  \bibinfo{author}{Q.~Wang},
\newblock \bibinfo{title}{An improved fourier series solution for free
  vibration analysis of the moderately thick laminated composite rectangular
  plate with non-uniform boundary conditions},
\newblock \bibinfo{journal}{International Journal of Mechanical Sciences}
  \bibinfo{volume}{121} (\bibinfo{year}{2017}) \bibinfo{pages}{1--20}.
  \DOIprefix\doi{10.1016/j.ijmecsci.2016.12.007}.
%Type = Article
\bibitem[{Senjanovic et~al.(2018)Senjanovic, Alujevic, Catipovic, Cakmak, and
  Vladimir}]{SENJANOVIC2018870}
\bibinfo{author}{I.~Senjanovic}, \bibinfo{author}{N.~Alujevic},
  \bibinfo{author}{I.~Catipovic}, \bibinfo{author}{D.~Cakmak},
  \bibinfo{author}{N.~Vladimir},
\newblock \bibinfo{title}{Vibration analysis of rotating toroidal shell by the
  rayleigh-ritz method and fourier series},
\newblock \bibinfo{journal}{Eng. Struct.} \bibinfo{volume}{173}
  (\bibinfo{year}{2018}) \bibinfo{pages}{870--891}.
  \DOIprefix\doi{10.1016/j.engstruct.2018.07.029}.
%Type = Article
\bibitem[{Barulich et~al.(2022)Barulich, Deutsch, Eisenberger, Godoy, and
  Dardati}]{Barulich2022}
\bibinfo{author}{N.~D. Barulich}, \bibinfo{author}{A.~Deutsch},
  \bibinfo{author}{M.~Eisenberger}, \bibinfo{author}{L.~A. Godoy},
  \bibinfo{author}{P.~M. Dardati},
\newblock \bibinfo{title}{A modified fourier series-based solution with
  improved rate of convergence for two-dimensional rectangular isotropic linear
  elastic solids},
\newblock \bibinfo{journal}{Math. Mech. Solids.} \bibinfo{volume}{27}
  (\bibinfo{year}{2022}) \bibinfo{pages}{410--432}.
  \DOIprefix\doi{10.1177/10812865211025584}.
%Type = Article
\bibitem[{Xu and Wu(2022)}]{Xu2022}
\bibinfo{author}{Y.~Xu}, \bibinfo{author}{Z.~Wu},
\newblock \bibinfo{title}{Exact solutions for rectangular anisotropic plates
  with four clamped edges},
\newblock \bibinfo{journal}{Mech. Adv. Mater. Struct.} \bibinfo{volume}{29}
  (\bibinfo{year}{2022}) \bibinfo{pages}{1756--1768}.
  \DOIprefix\doi{10.1080/15376494.2020.1838007}.
%Type = Article
\bibitem[{Nelisse et~al.(1998)Nelisse, Beslin, and Nicolas}]{Nelisse1998}
\bibinfo{author}{H.~Nelisse}, \bibinfo{author}{O.~Beslin},
  \bibinfo{author}{J.~Nicolas},
\newblock \bibinfo{title}{A generalized approach for the acoustic radiation
  from a baffled or unbaffled plate with arbitrary boundary conditions,
  immersed in a light or heavy fluid},
\newblock \bibinfo{journal}{J. Sound Vib.} \bibinfo{volume}{211}
  (\bibinfo{year}{1998}) \bibinfo{pages}{207--225}.
  \DOIprefix\doi{10.1006/jsvi.1997.1359}.
%Type = Article
\bibitem[{Zhang and Li(2010)}]{ZHANG20105307}
\bibinfo{author}{X.~Zhang}, \bibinfo{author}{W.~L. Li},
\newblock \bibinfo{title}{A unified approach for predicting sound radiation
  from baffled rectangular plates with arbitrary boundary conditions},
\newblock \bibinfo{journal}{J. Sound Vib.} \bibinfo{volume}{329}
  (\bibinfo{year}{2010}) \bibinfo{pages}{5307--5320}.
  \DOIprefix\doi{10.1016/j.jsv.2010.07.014}.
%Type = Article
\bibitem[{Zhang et~al.(2023)Zhang, Cui, Xu, and Peng}]{Zhang202313}
\bibinfo{author}{W.~W. Zhang}, \bibinfo{author}{L.~L. Cui},
  \bibinfo{author}{R.~W. Xu}, \bibinfo{author}{Y.~Peng},
\newblock \bibinfo{title}{{Fast analytical approximations for the acoustic
  radiation impedance of rectangular plates with arbitrary boundary
  conditions}},
\newblock \bibinfo{journal}{AIP Advances} \bibinfo{volume}{13}
  (\bibinfo{year}{2023}) \bibinfo{pages}{045116}.
  \DOIprefix\doi{10.1063/5.0131401}.
%Type = Article
\bibitem[{Zhao et~al.(2024)Zhao, Ye, Chen, Jin, Chen, and Liu}]{Zhao2024}
\bibinfo{author}{T.~Zhao}, \bibinfo{author}{T.~Ye}, \bibinfo{author}{Y.~Chen},
  \bibinfo{author}{G.~Jin}, \bibinfo{author}{Y.~Chen},
  \bibinfo{author}{Z.~Liu},
\newblock \bibinfo{title}{A fast chebyshev spectral approach for vibroacoustic
  behavior analysis of heavy fluid-loaded baffled rectangular plates with
  general boundary conditions},
\newblock \bibinfo{journal}{Thin-Walled Struct.} \bibinfo{volume}{196}
  (\bibinfo{year}{2024}) \bibinfo{pages}{111518}.
  \DOIprefix\doi{10.1016/j.tws.2023.111518}.
%Type = Article
\bibitem[{Zhang et~al.(2017)Zhang, Shi, and Wang}]{Zhang2017}
\bibinfo{author}{H.~Zhang}, \bibinfo{author}{D.~Shi},
  \bibinfo{author}{Q.~Wang},
\newblock \bibinfo{title}{An improved fourier series solution for free
  vibration analysis of the moderately thick laminated composite rectangular
  plate with non-uniform boundary conditions},
\newblock \bibinfo{journal}{Int. J. Mech. Sci.} \bibinfo{volume}{121}
  (\bibinfo{year}{2017}) \bibinfo{pages}{1--20}.
  \DOIprefix\doi{10.1016/j.ijmecsci.2016.12.007}.
%Type = Article
\bibitem[{Wang et~al.(2016)Wang, Shi, and Shi}]{Wang2016-2}
\bibinfo{author}{Q.~Wang}, \bibinfo{author}{D.~Shi}, \bibinfo{author}{X.~Shi},
\newblock \bibinfo{title}{A modified solution for the free vibration analysis
  of moderately thick orthotropic rectangular plates with general boundary
  conditions, internal line supports and resting on elastic foundation},
\newblock \bibinfo{journal}{Meccanica} \bibinfo{volume}{51}
  (\bibinfo{year}{2016}) \bibinfo{pages}{1985--2017}.
  \DOIprefix\doi{10.1007/s11012-015-0345-3}.
%Type = Inbook
\bibitem[{Stupakov and Penn(2018)}]{Stupakov2018}
\bibinfo{author}{G.~Stupakov}, \bibinfo{author}{G.~Penn}, \bibinfo{title}{The
  Basic Formulation of Mechanics: Lagrangian and Hamiltonian Equations of
  Motion}, \bibinfo{publisher}{Springer International Publishing},
  \bibinfo{address}{Cham}, \bibinfo{year}{2018}, pp. \bibinfo{pages}{3--19}.
  \DOIprefix\doi{10.1007/978-3-319-90188-6_1}.
%Type = Article
\bibitem[{Li(2004)}]{Li2004}
\bibinfo{author}{W.~L. Li},
\newblock \bibinfo{title}{Vibration analysis of rectangular plates with general
  elastic boundary supports},
\newblock \bibinfo{journal}{J. Sound Vib.} \bibinfo{volume}{273}
  (\bibinfo{year}{2004}) \bibinfo{pages}{619--635}.
  \DOIprefix\doi{10.1016/S0022-460X(03)00562-5}.
%Type = Inproceedings
\bibitem[{Deng et~al.(2020)Deng, Shao, Zheng, and Wu}]{Deng2020b}
\bibinfo{author}{G.~Deng}, \bibinfo{author}{J.~Shao},
  \bibinfo{author}{S.~Zheng}, \bibinfo{author}{X.~Wu},
\newblock \bibinfo{title}{Vibrational analysis of rectangular thin layer using
  fourier sine series as the auxiliary function},
\newblock in: \bibinfo{booktitle}{Proceedings of 2020 International Congress on
  Noise Control Engineering, INTER-NOISE 2020}, \bibinfo{address}{Seoul,
  Korea}, \bibinfo{year}{2020}.
%Type = Article
\bibitem[{Gao et~al.(2022)Gao, Pang, Li, and Jia}]{Gao2022}
\bibinfo{author}{C.~Gao}, \bibinfo{author}{F.~Pang}, \bibinfo{author}{H.~Li},
  \bibinfo{author}{D.~Jia},
\newblock \bibinfo{title}{A semi-analytical method for the dynamic
  characteristics of stiffened plate with general boundary conditions},
\newblock \bibinfo{journal}{Thin-Walled Struct.} \bibinfo{volume}{178}
  (\bibinfo{year}{2022}) \bibinfo{pages}{109513}.
  \DOIprefix\doi{10.1016/j.tws.2022.109513}.
%Type = Article
\bibitem[{Rhazi and Atalla(2010)}]{Rhazi2010}
\bibinfo{author}{D.~Rhazi}, \bibinfo{author}{N.~Atalla},
\newblock \bibinfo{title}{A simple method to account for size effects in the
  transfer matrix method},
\newblock \bibinfo{journal}{J. Acoust. Soc. Am.} \bibinfo{volume}{127}
  (\bibinfo{year}{2010}) \bibinfo{pages}{EL30--EL36}.
  \DOIprefix\doi{10.1121/1.3280237}.
%Type = Book
\bibitem[{Blevins(1979)}]{Blevins1979}
\bibinfo{author}{R.~D. Blevins}, \bibinfo{title}{Formulas for natural frequency
  and mode shape}, \bibinfo{publisher}{New York: Van Nostrand Reinhold
  Company}, \bibinfo{year}{1979}.
%Type = Article
\bibitem[{Zhang and Cheng(2017)}]{Zhang2017b}
\bibinfo{author}{S.~Zhang}, \bibinfo{author}{L.~Cheng},
\newblock \bibinfo{title}{Wavelet decompositions for high frequency vibrational
  analyses of plates},
\newblock \bibinfo{journal}{Int. J. Appl. Mech.} \bibinfo{volume}{09}
  (\bibinfo{year}{2017}) \bibinfo{pages}{1750088}.
  \DOIprefix\doi{10.1142/S1758825117500880}.
%Type = Article
\bibitem[{Ma et~al.(2018)Ma, Zhang, and Cheng}]{Ma2018}
\bibinfo{author}{L.~Ma}, \bibinfo{author}{S.~Zhang},
  \bibinfo{author}{L.~Cheng},
\newblock \bibinfo{title}{A 2d daubechies wavelet model on the vibration of
  rectangular plates containing strip indentations with a parabolic thickness
  profile},
\newblock \bibinfo{journal}{J. Sound Vib.} \bibinfo{volume}{429}
  (\bibinfo{year}{2018}) \bibinfo{pages}{130--146}.
  \DOIprefix\doi{10.1016/j.jsv.2018.04.042}.
%Type = Article
\bibitem[{Langley(2007)}]{Langley2007}
\bibinfo{author}{R.~S. Langley},
\newblock \bibinfo{title}{{On the diffuse field reciprocity relationship and
  vibrational energy variance in a random subsystem at high frequencies}},
\newblock \bibinfo{journal}{J. Acoust. Soc. Am.} \bibinfo{volume}{121}
  (\bibinfo{year}{2007}) \bibinfo{pages}{913--921}.
  \DOIprefix\doi{10.1121/1.2409484}.

\end{thebibliography}
\addcontentsline{toc}{section}{References}




%%\bibitem{}

%%\end{thebibliography}
\end{document}